\documentclass[10pt,a4paper]{article}
\usepackage{jheppub}

\usepackage[utf8]{inputenc}
\usepackage{float}
\usepackage{amsmath}
\usepackage[english]{babel}
\usepackage{amssymb}
\usepackage{mathtools}
\usepackage{color}
\definecolor{ao}{rgb}{0.0, 0.5, 0.0}

\newcommand{\avg}[1]{\langle #1 \rangle}

\usepackage{graphicx}
\usepackage{longtable}
\usepackage{tikz}
\usetikzlibrary{backgrounds}
\usepackage{comment}
\usepackage{capt-of}
\usepackage{hyperref}
\hypersetup{
    colorlinks=true,
    linkcolor=blue
    }

\usepackage{bm}

\usepackage{caption}
\usepackage{subcaption}

\newcommand{\mycomment}[1]{}
\newcommand{\Mod}[1]{\ (\mathrm{mod}\ #1)}

\def\z{z^{\dag}}
\def\w{w^{\dag}}
\def\s{s^{\dag}}

\def\pz{p^{\z}}
\def\ps{p^{\s}}
\def\pw{p^{\w}}
\def\pdagz{\bar{p}^{\z}}
\def\pdags{\bar{p}^{\s}}
\def\pdagw{\bar{p}^{\w}}
\def\qsigma{\bar{q}^{\sigma}}
\def\qphi{\bar{q}^{\phi}}
\def\qtau{\bar{q}^{\tau}}
\def\0{\mbox{\tiny $0$}}
\def\1{\mbox{\tiny $1$}}
\def\2{\mbox{\tiny $2$}}
\def\3{\mbox{\tiny $3$}}
\def\4{\mbox{\tiny $4$}}
\def\5{\mbox{\tiny $5$}}
\def\6{\mbox{\tiny $6$}}
\def\7{\mbox{\tiny $7$}}
\def\8{\mbox{\tiny $8$}}
\def\9{\mbox{\tiny $9$}}

\def\N{\bar{N}}

\def\llm{{\bar{m}^{\phi}_{\chi^1}}}
\def\m{{\bar{m}^{\sigma}_{\xi^1}}}
\def\n{{\bar{m}^{\tau}_{\eta^1}}}

\def\P{\bar{P}}
\def\q{\bar{q}}

\def\b{\beta^{'}}
\def\P{\mathbb{P}}

\DeclareMathOperator{\sign}{sign}
\DeclareMathOperator{\erf}{erf}
\def\a{g_{\sigma\tau}}
\def\b{g_{\sigma\phi}}
\def\c{g_{\tau\phi}}

\usepackage{tikz}
\usepackage{listofitems} 
\usetikzlibrary{arrows.meta} 
\usepackage[outline]{contour} 
\contourlength{1.4pt}

\tikzset{>=latex} 
\usepackage{xcolor}
\colorlet{myred}{red!80!black}
\colorlet{myblue}{blue!80!black}
\colorlet{mygreen}{green!60!black}
\colorlet{myyellow}{yellow!60!black}
\colorlet{myorange}{orange!85!red!90!black}
\colorlet{mydarkred}{red!30!black}
\colorlet{mydarkblue}{blue!40!black}
\colorlet{mydarkgreen}{green!30!black}
\tikzstyle{node}=[thick,circle,draw=myblue,minimum size=22,inner sep=0.5,outer sep=0.6]
\tikzstyle{node1}=[thick,rectangle,draw=myblue,minimum width = 0.7cm,  minimum height = 0.7cm,inner sep=-0.01,outer sep=0.01]
\tikzstyle{node in}=[node,green!20!black,draw=mygreen!30!black,fill=mygreen!25]
\tikzstyle{nodesigma}=[node,green!20!black,draw=mygreen!30!black,fill=mygreen!10]
\tikzstyle{nodetau}=[node,orange!20!black,draw=mygreen!30!black,fill=myorange!20]
\tikzstyle{nodephi}=[node,green!20!black,draw=mygreen!30!black,fill=myred!20]
\tikzstyle{node hidden}=[node1,blue!20!black,draw=myblue!30!black,fill=myblue!05]
\tikzstyle{node convol}=[node,orange!20!black,draw=myorange!30!black,fill=myorange!20]
\tikzstyle{node out}=[node,red!20!black,draw=myred!30!black,fill=myred!20]
\tikzstyle{connect}=[thick,mydarkblue] 
\tikzstyle{connect arrow}=[-{Latex[length=4,width=3.5]},thick,mydarkblue,shorten <=0.5,shorten >=1]
\tikzset{ 
  node 1/.style={node in},
  node 2/.style={node hidden},
  node 3/.style={node out},
}

\usetikzlibrary{fit}
\newcommand{\lr}[1]{\left(#1\right)}

\newcommand{\SOMMA}[2]{\displaystyle\sum\limits_{#1}^{#2}}

\long\def \beq#1\eeq {\begin{equation} #1 \end{equation}}
\long\def \beaq#1\eeaq {\begin{equation}\begin{aligned} #1 \end{aligned}\end{equation}}
\long\def \bes#1\ees {\begin{equation}\begin{split} #1 \end{split} \end{equation}}
\long\def \bea#1\eea {\begin{eqnarray} #1 \end{eqnarray}}
\long\def \bse[#1]#2\ese {\begin{subequations}\label{#1}\begin{align} #2 \end{align}\end{subequations}}

\title{Generalized hetero-associative neural networks}
\author[a]{Elena Agliari,}
\author[b]{Andrea Alessandrelli,}
\author[c,d]{Adriano Barra,}
\author[e]{Martino Salomone Centonze,}
\author[f]{Federico Ricci-Tersenghi}

\affiliation[a]{Dipartimento di Matematica, Sapienza Universit\`a di Roma, Rome, Italy.}

\affiliation[b]{Dipartimento di Informatica, Universit\`a di Pisa, Pisa Italy.}

\affiliation[c]{Dipartimento di Scienze di Base ed Applicate per l’Ingegneria, Sapienza   Universit\`a  di Roma, Rome, Italy}    

\affiliation[d]{Istituto Nazionale di Fisica Nucleare, Sezioni di Lecce $\&$ Roma, Italy.}

\affiliation[e]{Dipartimento di Matematica, Universit\`a di Bologna, Italy.}

\affiliation[f]{Dipartimento di Fisica, Sapienza Universit\`a di Roma, Rome, Italy.}

\abstract{Auto-associative neural networks (e.g., the Hopfield model implementing the standard Hebbian prescription) serve as a foundational framework for pattern recognition and associative memory in statistical mechanics. However, their hetero-associative counterparts, though less explored, exhibit even richer computational capabilities.
In this work, we examine a straightforward extension of Kosko's Bidirectional Associative Memory (BAM), introducing a Three-directional Associative Memory (TAM), that is a tripartite neural network equipped with generalized Hebbian weights.
Through both analytical approaches (using replica-symmetric statistical mechanics) and computational methods (via Monte Carlo simulations), we derive phase diagrams within the space of control parameters, revealing a region where the network can successfully perform pattern recognition as well as other tasks tasks. In particular, it can achieve pattern disentanglement, namely, when presented with a mixture of patterns, the network can recover the original patterns. Furthermore, the system is capable of retrieving Markovian sequences of patterns and performing generalized frequency modulation.
}
  
\begin{document}

\maketitle
\flushbottom

\newpage
\section{Introduction}

Hetero-associative memories were proposed by Kosko in the eighties \cite{Kosko} as generalizations of the auto-associative ones \cite{AGS1,AGS2,Amit,Coolen}:
the so-called {\em bidirectional associative memories} (BAM) displays neurons that are split into two layers and the inter-layer interactions are defined as generalized Hebbian couplings that combine patterns belonging to two different classes \cite{Luca,Brunello,BAM-noi1,BAM-noi2,Taro}. As a result of these different designs, auto-associative and bi-layer hetero-associative networks perform different kinds of task: if properly stimulated, the former can retrieve single patterns sequentially, while the latter can simultaneously retrieve pairwise-associated patterns. 
So far, the analysis of hetero-associative networks 
has been confined to architectures equipped with two layers and here we extend the analysis to three-layer networks, referred to as three-directional associative memories (TAM); we shall as well extend the dataset to encompass three classes of patterns. We conduct a comprehensive statistical mechanical analysis of these networks, at the replica symmetric level of description \cite{MPV,Amit,Fachechi1,Guerra2,Huang}, whose findings are corroborated  by numerics (mainly Monte Carlo simulations) to prove that these networks show richer spontaneous information processing capabilities, if compared to the auto-associative counterpart and to the BAM. 
\newline
In particular, as mentioned above, in this setting classic {\em pattern recognition} task becomes a {\em generalized pattern recognition} task  because, when supplied with a noisy pattern on one of its layers, this network retrieves that pattern on that layer as well as associated patterns on the other two layers. Thus, perceiving a noisy pattern can now trigger the retrieval of a bunch of memories, mirroring {\em generalized conditioning} in Pavlovian learning \cite{Aquaro-Pavlov}.  
\newline
Moreover, when fed with a signal that is a combination of patterns, this network can recognize and separate all the patterns giving rise to the mixture. More precisely, a mixture of patterns is supplied as input and the constituting patterns are retrieved separately and simultaneously by different layers. We refer to this task as {\em pattern disentanglement} and we emphasize that there is no counterpart of this task in auto-associative memories.
\newline
We can further enrich this picture by moving from statically-retrieved patterns to dynamically-retrieved patterns: TAMs are capable of reconstructing temporal sequences of patterns and even combinations of sequences, regardless of whether the patterns occur at the same frequency or at different frequencies. In the latter setting, by using the slowly evolving sequence of patterns as the base-band, TAMs can  perform \emph{generalized frequency modulation}. 
\newline
Before deepening the aforementioned applications, our primary goal is to provide a statistical mechanical picture of the TAM so to obtain a macroscopic description of the system as a function of its control parameters, possibly summarized in terms of a phase diagram. 
There are plenty of techniques to reach this goal (see e.g. \cite{Lenka,Pedreschi1,Pedreschi2,Riccardo,Barbier,Bovier,Carleo,Fede-Data2014,FedeNew,AuroRev}) and, in this paper, we rely on Guerra's interpolation approach under the self-averaging assumption for the order parameters (i.e. at the replica symmetric level of description \cite{MPV,Coolen}, fairly standard in neural network theory \cite{Amit,Nishimori,HuangBook}). 
\newline
We present the main theoretical results of this investigation in Sec.~\ref{Sec:Theory}, while we left Sec.~\ref{Sec:Applications} for the applications. More precisely, Sec.~\ref{Sec:Theory} is split into two main subsections: Sec.~\ref{cicciobomba}, where the neural network is introduced, and Sec. \ref{Guerra}, where  we provide a quantitative picture of its computational capabilities.
%
Next, in Sec.~\ref{Sec:Applications} we discuss and investigate applications: in Sec.~\ref{Sec:Application-1} we generalize the classic task of pattern recognition; in Sec. \ref{Sec:Application-2}, we face the new task of (static) pattern disentanglement by focusing on ``heterogeneous'' mixtures; in Sec. \ref{Sec:Penalty} we consider a more challenging case, where mixtures are ``homogeneous''; in Sec.~\ref{Sec:Application-3} we deal with a dynamical problem: given a temporal sequence of patterns, where single patterns appear with the same frequency, the network must recognize all of them, as well as their temporal order of appearance (for this task to be fulfilled the network has to know the oscillation frequency); in Sec. \ref{Sec:Application-4} we ask the network to deal with a more complex task that stems from the two previous points: we prepare two sequences of patterns characterized by two different frequencies and we generate a mixture that combines the two sequences; next, this mixture is provided as a signal and the networks is asked to disentangle this input and return the time-ordered series of patterns building up the mixture -- note that this task can be seen as a (structured or {\em encrypted}) generalization of standard {\em frequency modulation} simply by thinking at the slow series as the base-band signal and the fast one being the information carrier\footnote{We can think at this task as an encrypted frequency modulation because, should this TAM be used to decode the signal, it has to know -in advance- the ensemble of patterns to deal with,  such that another network -without the same weights- would not be able to split the base-band signal from the information carrier one.}. 
Finally, in Sec.~\ref{sec:conclusions} we summarize our results and possible outlooks, and in Apps.~\ref{Appendix-Guerra}-\ref{app:B} we collect the technical details underlying the analytical investigations.

\section{Three-directional associative memory (TAM): theory}\label{Sec:Theory}
\subsection{Definitions}\label{cicciobomba}
We propose a tripartite generalization of Kosko's BAM \cite{Kosko} -- referred to as Three-directional Associative Memory (TAM) --  where the neural layers, hereafter indicated by $\bm\sigma \equiv \{\sigma_i\}_{i=1,..,N}$,  $\bm\tau \equiv  \{\tau_i\}_{i=1,..,M}$ and $\bm\phi \equiv \{\phi_i\}_{i=1,..,L}$, interact in pairs via generalized Hebbian couplings (\emph{vide infra}) built on $K$ triplets of binary patterns, \emph{i.e.} $\{(\bm\xi^\mu, \bm\eta^\mu, \bm\chi^\mu)\}_{\mu=1,..,K}$. Specifically, $\boldsymbol \xi^{\mu} \in \{-1, +1\}^N$, $\boldsymbol \eta^{\mu} \in \{-1, +1\}^M$, $\boldsymbol \chi^{\mu} \in \{-1, +1\}^L$ and their entries are all taken as i.i.d. Rademacher random variables.

\begin{figure}[tb]
    \centering, 
    \includegraphics[width=7cm]{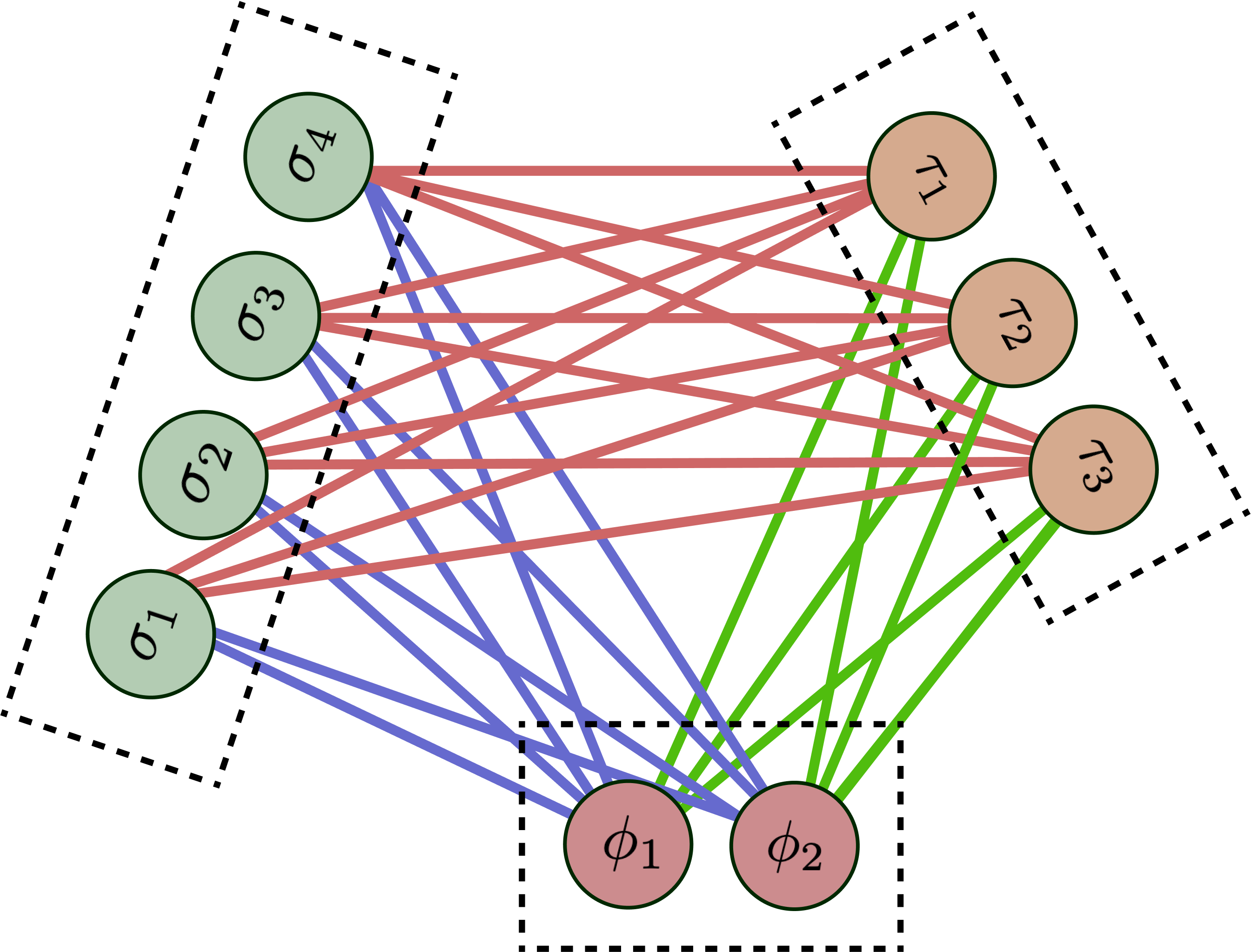}
    \caption{Representation of the TAM neural network: each layer is built of by binary (Ising) units that interact pair-wise in a generalized Hebbian fashion, see the synaptic matrices in eq.s \eqref{Sinapsi1}-\eqref{Sinapsi3} plugged into the cost function \eqref{H-TAM-riscritta}. In particular, in this example, the layer $\sigma$ is built of by $N=4$ neurons, the layer $\tau$ by $M=3$ neurons and the layer $\phi$ by $L=2$ neurons. The red links pertain to $\bm J^{\sigma \tau}$, the blue ones to $\bm J^{\sigma \phi}$, and the green ones to $\bm J^{\tau \phi}$.}
    \label{fig:TAM-diretta}
\end{figure}

The Hamiltonian (or {\em cost function}) of the TAM model reads as
\begin{equation}\label{H-TAM-diretta}\small
    \mathcal{H}_{N,M,L}(\bm\sigma,\bm\tau,\bm\phi|\bm\xi,\bm\eta,\bm\chi)=-\SOMMA{\mu=1}{K}\left(\dfrac{\a}{\sqrt{NM}}\SOMMA{i,j=1}{N,M}\xi_i^\mu\eta_j^\mu\sigma_i\tau_j+\dfrac{\b}{\sqrt{NL}}\SOMMA{i,j=1}{N,L}\xi_i^\mu\chi_j^\mu\sigma_i\phi_j+\dfrac{\c}{\sqrt{M L}}\SOMMA{i,j=1}{L,M}\chi_i^\mu\eta_j^\mu\phi_i\tau_j\right)\, ,
\end{equation}
where $g=(\a,\b,\c)\in \mathbb{R}^3$ tune the relative strength of inter-layer interactions while the presence of the factors $N,M,L$ in the denominators ensures a correct scaling of the Hamiltonian in the thermodynamic limit $N, M, L \to \infty$, where our calculations will be carried out; a sketch of the TAM architecture is shown in Fig.~\ref{fig:TAM-diretta}. 
Note that the $\sigma$-layer interacts with the $\tau$-layer via a synaptic matrix $\bm J^{\sigma \tau}$ with entries
\begin{equation}\label{Sinapsi1}
J_{ij}^{\sigma \tau} = \dfrac{1}{\sqrt{NM}}\SOMMA{\mu=1}{K}\xi_i^\mu\eta_j^\mu, 
\end{equation}
and with the $\phi$-layer via a synaptic matrix $\bm J^{\sigma \phi}$ with entries
\begin{equation}\label{Sinapsi2}
J_{ij}^{\sigma \phi} = \dfrac{1}{\sqrt{NL}}\SOMMA{\mu=1}{K}\xi_i^\mu\chi_j^\mu, 
\end{equation}
while the $\tau$-layer interacts with the $\phi$-layer via a synaptic matrix $\bm J^{\tau \phi}$ with entries
\begin{equation}\label{Sinapsi3}
J_{ij}^{\tau \phi} = \dfrac{1}{\sqrt{ML}}\SOMMA{\mu=1}{K}\eta_i^\mu\chi_j^\mu,
\end{equation}
such that the cost function \eqref{H-TAM-diretta} can be rewritten as
\begin{equation}\label{H-TAM-riscritta}
 \mathcal{H}_{N,M,L}(\bm\sigma,\bm\tau,\bm\phi|\bm\xi,\bm\eta,\bm\chi)= - \left(\a \SOMMA{i,j=1}{N,M}  J^{\sigma \tau}_{ij} \sigma_i \tau_j + \b
\SOMMA{i,j=1}{M,L}  J^{\sigma \phi}_{ij} \sigma_i \phi_j + \c
\SOMMA{i,j=1}{N,L}  J^{\tau \phi}_{ij} \tau_i \phi_j\right),
\end{equation}
where it shines that the layer $\boldsymbol \sigma$ (respectively, layer $\boldsymbol\tau$ and $\boldsymbol \phi$) is responsible for the $\boldsymbol\xi$ dataset (respectively, the $\boldsymbol \eta$ and $\boldsymbol\chi$ datasets). The equations \eqref{Sinapsi1}, \eqref{Sinapsi2}, \eqref{Sinapsi3} define the so-called generalized hetero-associative Hebbian couplings \cite{Kosko}. 

We stress that the size of the three layers and, accordingly, the length of the related patterns, can be different, that is $N \neq M \neq L$, nevertheless, their amount must be  the same, that is $K$ patterns per each layer. Furthermore, the ratio between the amount of patterns and their length must stay finite in the asymptotic (thermodynamic) limit, namely, we choose $K,N,M,L$ such that
\begin{equation}
\begin{array}{lllllll}
      \lim\limits_{N,L\to\infty}\sqrt{\dfrac{N}{L}}=\alpha\,,\; && \lim\limits_{M,N\to\infty}\sqrt{\dfrac{N}{M}}=\theta\,,\;&&\lim\limits_{M,L\to\infty}\sqrt{\dfrac{M}{L}}=\dfrac{\alpha}{\theta} \,,\; &&\lim\limits_{N\to\infty}\dfrac{K}{N}=\gamma\,,
\end{array}
\end{equation}
with $\alpha,\theta,\gamma\in\mathbb{R}^+$. Thus, $\gamma$ defines the storage of the network and we will study both the high-storage regime ($\gamma > 0$) and the low-storage regime ($\gamma \to 0$).

Further, denoted by $\mathcal{P}_{t}(\bm\sigma,\bm\tau,\bm\phi|\bm\xi,\bm\eta,\bm\chi)$ the probability of finding the system in a configuration $(\bm\sigma,\bm\tau,\bm\phi)$ at time $t$, the following master equation
\begin{equation}\label{dinamical}
\mathcal{P}_{t+1}(\bm\sigma,\bm\tau,\bm\phi|\bm\xi,\bm\eta,\bm\chi) = \sum_{(\bm \sigma',\bm \tau', \bm\phi') \in \{-1, +1\}^{N+M+L} } W_{\beta}(\bm\sigma',\bm\tau',\bm\phi' \to \bm\sigma,\bm\tau,\bm\phi|\bm\xi,\bm\eta,\bm\chi)\mathcal{P}_t(\bm \sigma',\bm \tau',\bm \phi'\vert \bm \xi, \bm \eta,\bm \chi)
\end{equation}
describes a Markov process in the space of neural configurations, where $W_{\beta}$  represents the transition rate from a state $(\bm\sigma',\bm\tau',\bm\phi')$ to a state $(\bm\sigma,\bm\tau,\bm\phi)$ and it is chosen in such a way that the system is likely to lower the value of the cost function \eqref{H-TAM-diretta} along its evolution: this likelihood is tuned by the parameter $\beta \in \mathbb R^+$ such that for $\beta \to 0^+$ the dynamics is a pure  random walk in the neural configuration space (and any configuration is equally likely to occur), while for $\beta \to +\infty$ the dynamics steepest descends toward the minima of the Hamiltonian \cite{Coolen}. Remarkably, the symmetry of the pairwise couplings in the TAM cost function is enough for detailed balance to hold 
and, with a suitable choice for $W_{\beta}$, the long-time limit of the stochastic process \eqref{dinamical} relaxes to the Boltzmann-Gibbs distribution 
\begin{equation}\label{BGmeasure}
\lim_{t \to \infty}\mathcal{P}_t(\bm\sigma,\bm\tau,\bm\phi|\bm\xi,\bm\eta,\bm\chi) = \mathcal{P}(\bm\sigma,\bm\tau,\bm\phi|\bm\xi,\bm\eta,\bm\chi) = \frac{e^{-\beta \mathcal{H}_{N,L,M}(\bm\sigma,\bm\tau,\bm\phi|\bm\xi,\bm\eta,\bm\chi)}}{\mathcal Z_{N,M,L}(\beta)}
\end{equation}
where the normalization factor $\mathcal Z_{N,M,L}(\beta)$ reads as  
 \begin{equation}\label{partition-function}
    \mathcal Z_{N,M,L}(\beta) = \sum_{(\bm \sigma,\bm \tau, \bm\phi) \in \{-1, +1\}^{N+M+L} } e^{-\beta \mathcal{H}_{N,M,L}(\bm\sigma,\bm\tau,\bm\phi|\bm\xi,\bm\eta,\bm\chi)},
\end{equation}
and is also referred to as {\em partition function}; the system's emergent behavior can thus be analyzed with tools borrowed from statistical mechanics.

\subsection{Statistical mechanics investigation}\label{Guerra}
Pivotal for a statistical mechanical analysis is the study of the asymptotic\footnote{This means that we are working in the limit of infinite size; this condition makes the analytical treatment feasible. It is worth recalling that, order to preserve $(\alpha, \theta) \in \mathbb{R}^+$, the infinite volume limit $N\to \infty$ implies suitably divergences also of the sizes of the two other layers, i.e. $M$ and $L$. One may question why, to keep the logarithm intensive, we divided by $N$ rather than, say, $M$ or $L$ or their combinations, as e.g. $\sqrt{NM}$, etc. In fact, all these possible alternative definitions work as well as they just trivially re-scale outcomes obtained by using $N$ as we do here.} quenched free energy, defined as 
\begin{equation}\label{eq:Free-Definition}
\mathcal A_{\alpha,\theta,\gamma}^{g}(\beta) = \lim_{N \to \infty}\mathbb{E}\frac{1}{N}\ln \mathcal Z_{N,M,L}(\beta), 
\end{equation}
where $\mathbb{E}$ averages over the pattern distribution.
\newline
More precisely, we aim to find an expression of $\mathcal A_{\alpha,\theta,\gamma}^{g}(\beta)$, depending on a suitable set of macroscopic observables able to describe the emerging behavior of the system. As standard for these Hopfield-like networks, we need two sets of observables that assess, respectively, the quality of retrieval and the extent of frustration yielding to glassy phenomena. Indeed, we introduce the Mattis magnetizations 
\begin{eqnarray} \label{eq:ma}
         m_{\xi^\mu}^\sigma &=& \dfrac{1}{N}\SOMMA{i=1}{N}\xi_i^\mu\sigma_i,\\  m_{\eta^\mu}^\tau &=& \dfrac{1}{M}\SOMMA{i=1}{M}\eta_i^\mu\tau_i,\label{eq:ma_t}\\
         m_{\chi^\mu}^\phi &=& \dfrac{1}{L}\SOMMA{i=1}{L}\chi_i^\mu\phi_i,\label{eq:ma_p}
\end{eqnarray}
and the replica overlaps
\begin{eqnarray}
         q^{\sigma}_{ab} = \dfrac{1}{N}\SOMMA{i=1}{N}\sigma_i^{(a)}\sigma_i^{(b)},\\ q^\tau_{ab} = \dfrac{1}{M}\SOMMA{i=1}{M}\tau_i^{(a)}\tau_i^{(b)}, \\
         \label{eq:qc}
         q^\phi_{ab} = \dfrac{1}{L}\SOMMA{i=1}{L}\phi_i^{(a)}\phi_i^{(b)},
  \end{eqnarray}       
where $a$ and $b$ label two different replicas. The Mattis magnetizations and the overlaps are also referred to as \emph{order parameters}.

\begin{figure}[t]
    \centering
    \includegraphics[width=10cm]{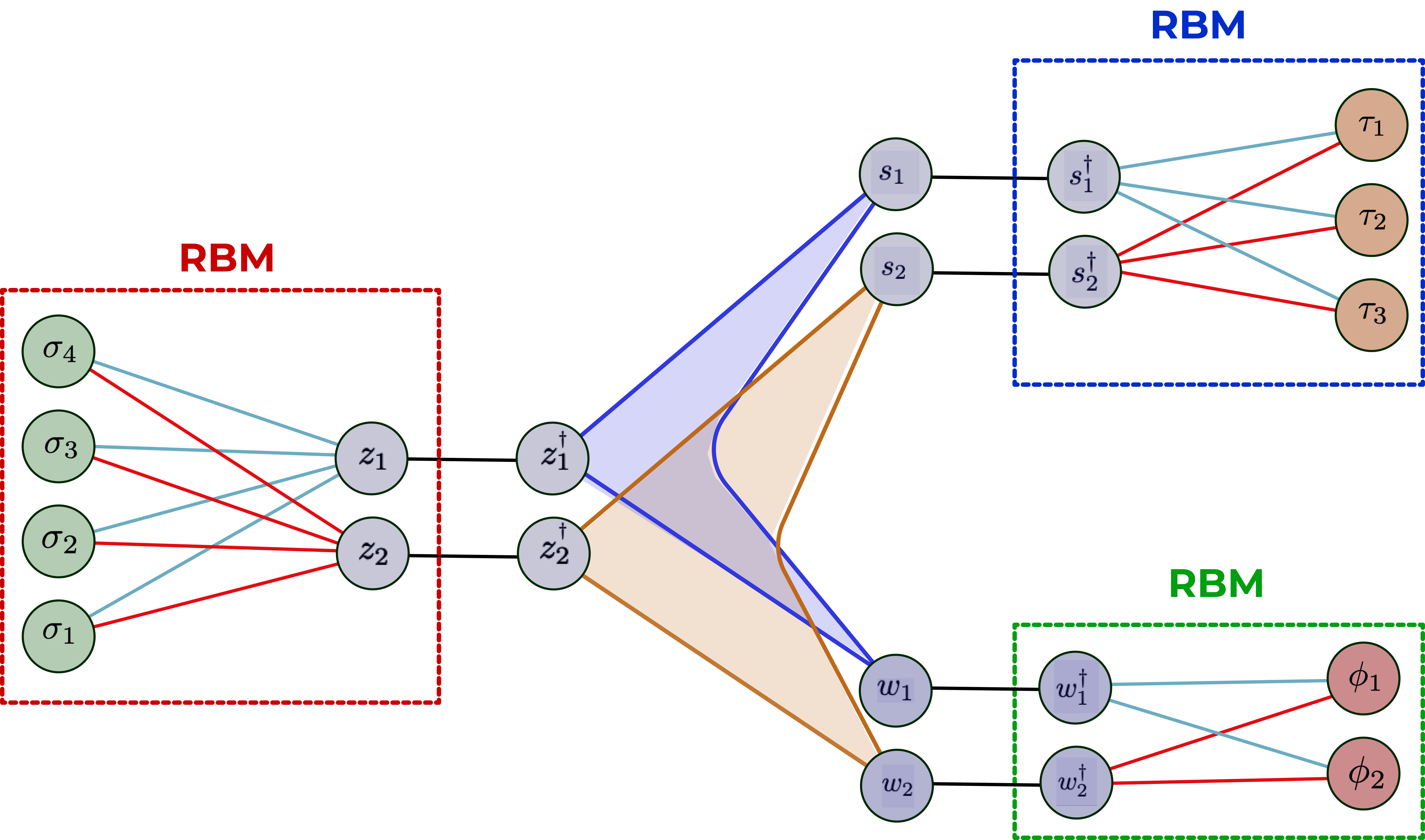}
    \caption{Equivalent representation of the TAM neural network shown in Fig. \ref{fig:TAM-diretta} as suggested by Eq. \eqref{eq:rappresentazioneintegrale}. The three visible layers are no longer directly interacting, but each visible layer is connected to a hidden layer which, in turn, is connected to a conjugated hidden layer. The visible layer is built of by standard (i.e., not selective) binary neurons, while the hidden layers are built of by highly-selective grandmother units firing solely when the pattern they code for is presented, or reconstructed, on the corresponding visible layer. Furthermore, a visible layer and the adjacent hidden layer make up a restricted Boltzmann machine (RBM). Thus, we have three RBMs whose coupling is ruled by the internal conjugated hidden neurons: their interaction encodes for the association among the three datasets.}
    \label{fig:int_rapp_TAM}
\end{figure}

Next, exploiting the fact that the Hamiltonian is a quadratic form in the magnetizations and the Boltzmann-Gibbs measure is (proportional to) the exponential of the Hamiltonian, the partition function \eqref{partition-function} admits an  integral representation that reads as  
\begin{equation}\label{eq:rappresentazioneintegrale}
    \begin{array}{lllll}
         \mathcal Z_{N,M,L}(\beta)= \sum _{(\bm \sigma',\bm \tau', \bm\phi') \in \{-1, +1\}^{N+M+L} } \exp\left[\beta\left(\a\sqrt{NM} m_{\xi^1}^\sigma m_{\xi^1}^\tau+\b\sqrt{NL} m_{\xi^1}^\sigma m_{\xi^1}^\phi+\c\sqrt{LM}m_{\xi^1}^\phi m_{\xi^1}^\tau\right)\right]\times
         \\\\
         \times\displaystyle\int\mathcal{D}(z\z s\s w\w)\exp\left[\sqrt{\dfrac{\beta}{2N}}\SOMMA{\mu>1}{K}\SOMMA{i=1}{N}\xi_i^\mu\sigma_i z_\mu+\sqrt{\dfrac{\beta}{2M}}\SOMMA{\mu>1}{K}\SOMMA{j=1}{M}\eta_j^\mu\tau_j \s_\mu+\sqrt{\dfrac{\beta}{2\N}}\SOMMA{\mu>1}{K}\SOMMA{k=1}{L}\chi_k^\mu\phi_k \w_\mu\right]
    \end{array}
\end{equation}
where 
\begin{equation}
\label{Giacobbiani}
    \begin{array}{lllll}
         \displaystyle\int\mathcal{D}(z\z s\s w\w)=\prod\limits_{\mu>1}^K\displaystyle\int \left(\dfrac{dz_\mu b\z_\mu ds_\mu b\s_\mu dw_\mu b\w_\mu}{(\sqrt{2\pi})^6}\right)e^{\left[-\dfrac{1}{2}\left(z_\mu \z_\mu +s_\mu \s_\mu+w_\mu \w_\mu- \a\z_\mu s_\mu- \b\z_\mu w_\mu- \c s_\mu w_\mu\right)\right]}
    \end{array}
\end{equation}
and we have used the relation
\begin{equation}\label{GeneralGauss}
    \begin{array}{lllll}
         \exp\left[\beta\SOMMA{\mu>1}{K}A_\mu B_\mu \right]=\prod\limits_{\mu>1}^K\displaystyle\int \left(\dfrac{dx_\mu dx^\dag_\mu}{(\sqrt{2\pi})^2}\right)\exp\left[-\dfrac{1}{2}x_\mu x^\dag_\mu  + \sqrt{\dfrac{\beta}{2}} A_\mu x_\mu+ \sqrt{\dfrac{\beta}{2}}B_\mu x^\dag_\mu\right]\,
    \end{array}
\end{equation}
valid for any $A_{\mu}, B_{\mu} \in \mathbb R$. 
\newline
If we read the dummy integration variables, appearing in the exponent at the r.h.s. of eq.s \eqref{eq:rappresentazioneintegrale} and \eqref{Giacobbiani}, 
as hidden neurons, we notice that $\mathcal Z_{N,M,L}(\beta)$ corresponds to the partition function of the network sketched in Fig. \ref{fig:int_rapp_TAM}, which therefore constitutes an alternative, equivalent representation of the TAM. In particular, this exhibits six hidden layers, made of $K$ neurons each\footnote{This is a consequence of the identity \eqref{GeneralGauss} that implies that in each hidden layer there are exactly $K$ hidden neurons, one per stored pattern.} and these play as highly-selective neurons, namely units that, in the low-noise regime, become active if and only if the corresponding visible layer is aligned with a given pattern, say $\bm \xi^1$; for this reason they are also referred to as \emph{grandmother neurons} \cite{AABD-NN22}. Then, the activation of the hidden neuron $z_1$ prompts the activation of its conjugate $z_1^{\dag}$, which is directly connected with its label-mates $s_1$ and $w_1$, whose activation will in turn trigger the retrieval of the patterns $\bm \eta^1$ and $\bm \chi^1$ on the related visible layers. This explains why the TAM can handle triplets of patterns at once and the choice for the name of the task as {\em generalized pattern recognition}. We emphasize that, for the random, structureless database under study, hidden neurons capture the label of the various inputted patterns as this is the solely distinguishing feature, yet on structured datasets a single feature, and therefore a hidden neuron, may not be sufficient to unambiguously detect a pattern. 
\newline
As detailed in App.~\ref{Appendix-Guerra}, the linearized expression of the partition function allows us to exploit Guerra's interpolation method and reach an explicit expression for the quenched free energy; in our calculations we will assume replica symmetry, which implies that, in the thermodynamic limit, the observables \eqref{eq:ma}-\eqref{eq:qc} display vanishing fluctuations around their thermal average, hereafter denoted by a bar.
Once achieved an expression for the quenched free energy in terms of control and order parameters, we can extremize it w.r.t. the order parameters obtaining a set of self-consistent equations, whose solution provides the behavior of the order parameters as a function of the control parameters. The analysis of these solutions finally allows us to paint the phase diagram, highlighting the regions in the space of the control parameters where we anticipate a correct retrieval by the network.

\begin{figure}[t]
    \centering
     \includegraphics[height=4.5cm]{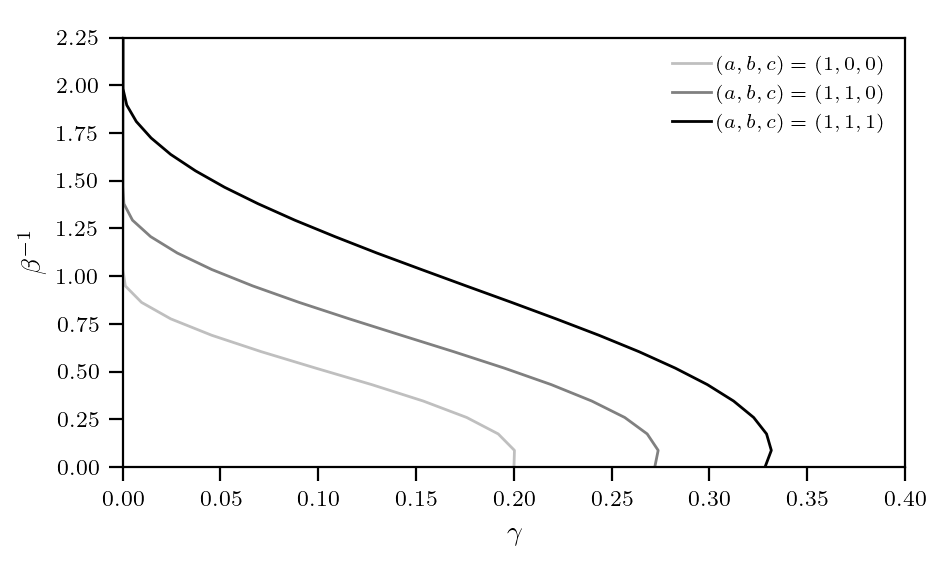}
     \includegraphics[height=4.5cm]{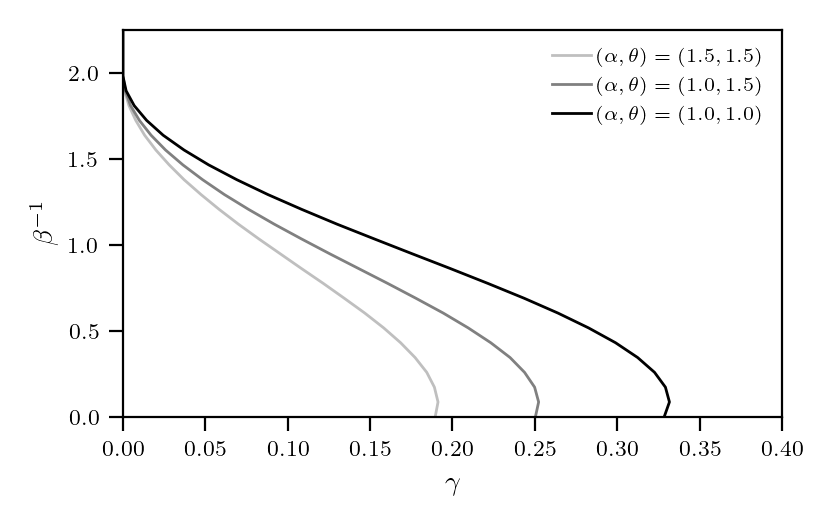}
    \caption{Left: Phase diagrams for the TAM model in the noise versus storage plane. Here we set $\alpha=\theta=1$ -- that is, the sizes of the various layers are set equal -- and we consider different inter-layer interaction strengths are considered as reported in the legend. The solid lines represent the boundary of the retrieval region, 
    obtained by asking for $|\m| , |\n|, |\llm| >0$ in the self-consistency equations \eqref{eq:self}: these inequalities hold simultaneously in the region below the solid line, while above the solid line the magnetizations are all vanishing. Note that the largest retrieval region is achieved for $(\a,\b,\c,)=(1,1,1)$, as expected given the relative abundance of links in this setting. Note further that the case $(\a,\b,\c,)=(1,0,0)$ correctly returns the BAM retrieval region \cite{BAM-noi1}. Right: Phase diagrams for the TAM model in the noise versus storage plane. Here we fix $(\a,\b,\c,)=(1,1,1)$ and we tune $(\alpha,\theta)$. Again, as intuitive, the largest retrieval region is obtained for the symmetric case, where $\alpha=\theta=1$ that is the configuration with more links.
    }
    \label{fig:TAM-PD}
\end{figure}

Focusing on the retrieval of the triplet of patterns with label $\mu=1$, without loss of generality, the self-consistency equations for the order parameters read as 
\begin{equation}
\label{eq:self}
    \begin{array}{lll}
         \m &=&\mathbb E_x\left\{\tanh\left[ \beta\n\dfrac{\a}{\theta}+ \beta\llm\dfrac{\b}{\alpha}+ x \sqrt{\beta\gamma\bar{p}^z/2}\right]\right\},
         \\\\
         \n &=&\mathbb E_x\left\{\tanh\left[\theta\left( \beta \a\m+ \beta\llm\dfrac{\c}{\alpha}+ x \sqrt{\beta\gamma\pdags/2}\right)\right]\right\},
         \\\\
         \llm &=&\mathbb E_x\left\{\tanh\left[\alpha\left( \beta \b\m+\beta\n\dfrac{\c}{\theta}+ x \sqrt{\beta\gamma\pdagw/2}\right)\right]\right\},
         \\\\
         \qsigma &=& \mathbb E_x\left\{\tanh^2\left[ \beta\n\dfrac{\a}{\theta}+ \beta\llm\dfrac{\b}{\alpha}+ x \sqrt{\beta\gamma\bar{p}^z/2}\right]\right\},
         \\\\
         \qtau &=& \mathbb E_x\left\{\tanh^2\left[\theta\left( \beta \a\m+ \beta\llm\dfrac{\c}{\alpha}+ x \sqrt{\beta\gamma\pdags/2}\right)\right]\right\},
         \\\\
         \qphi &=& \mathbb E_x\left\{\tanh^2\left[\alpha\left( \beta \b\m+\beta\n\dfrac{\c}{\theta}+ x \sqrt{\beta\gamma\pdagw/2}\right)\right]\right\},
    \end{array}
\end{equation}
where 
\begin{equation}
\nonumber
\begin{array}{lll}
\dfrac{\beta}{2}\gamma\bar{p}^z(\a,\b,\c,\beta,\qsigma,\qtau,\qphi)&=& 2\dfrac{\mathcal{D}(\qsigma,\qtau,\qphi) }{[\mathrm{det} \bm C]^2}\partial_{\qsigma}(\mathrm{det} \bm C)+2\dfrac{\gamma}{2}\dfrac{\partial_{\qsigma}(\mathrm{det} \bm C)}{\mathrm{det} \bm C}-2\dfrac{\partial_{\qsigma}(\mathcal{D}(\qsigma,\qtau,\qphi)) }{\mathrm{det} \bm C}
\end{array}
\end{equation}
(the explicit expression for the matrix $\bm C \in \mathbb C^{6 \times 6}$ is reported in App.~\ref{sec:Cauchy}, see eq.~\ref{eq:A14}) and 
\begin{equation} \nonumber
\begin{array}{lll}
     \mathcal{D}(\qsigma,\qtau,\qphi) &=& \gamma\beta^3\a\b\c\left[  \qsigma(1 - \qtau)(1 - \qphi)     +(1 - \qsigma) (1 - \qtau)\qphi    + (1 - \qsigma) \qtau   (1 - \qphi) \right]
    \\\\
    &&
    +\dfrac{\gamma}{2}\left[\beta^2 \a^2  (1 - \qsigma) \qtau + \beta^2 \a^2 \qsigma(1 - \qtau)  + 
 \beta^2 \c^2 \qtau(1 - \qphi )  \right]
 \\\\
 && +\dfrac{\gamma}{2}\left[\beta^2 \c^2 (1 - \qtau) \qphi + 
 \beta^2 \b^2 (1 - \qphi) \qsigma + \beta^2 \b^2(1 - \qsigma) \qphi\right].
\end{array}
\end{equation}
The self-consistency equations are solved numerically and the phase diagram of the model is obtained as a function of the coupling parameters $(\a,\b,\c)$ and of the relative sizes $(\alpha,\theta)$, see Fig. \ref{fig:TAM-PD}.
\newline
It is also instructive to evaluate explicitly, in the low storage limit $\gamma \to 0$, the critical noise that the model can tolerate before its information processing capabilities get lost. 
Starting from the self-consistency equations \eqref{eq:self}, we set  $\gamma=0$ to obtain 
\begin{equation} 
    \begin{array}{lll} 
        \begin{cases}
            \m =&\tanh\left[ \beta\n\dfrac{\a}{\theta}+ \beta\llm\dfrac{\b}{\alpha}\right] 
            \\\\ 
            \n =&\tanh\left[\theta\left( \beta \a\m+ \beta\llm\dfrac{\c}{\alpha}\right)\right] 
            \\\\ 
            \llm =&\tanh\left[\alpha\left( \beta \b\m+\beta\n\dfrac{\c}{\theta}\right)\right]
        \end{cases}
    \end{array} 
\end{equation}

Then, we linearize these equations near $(\m, \n,\llm)\sim(0,0,0)$\footnote{We exploit the Taylor expansion of the hyperbolic tangent function: $$\tanh(x)\underset{|x|\ll 1}{=}x-\frac{1}{3}x^3+\mathcal{O}(x^5),$$.}: 
\begin{equation} 
    \begin{array}{lll} 
        \begin{cases} 
        \m =& \beta\left(\n\dfrac{\a}{\theta}+ \llm\dfrac{\b}{\alpha}\right)-\dfrac{\beta^3}{3}\left( \n\dfrac{\a}{\theta}+ \llm\dfrac{\b}{\alpha}\right)^3 
        \\\\ 
        \n =&\beta\theta\left( \a\m+ \llm\dfrac{\c}{\alpha}\right)-\dfrac{\beta^3\theta^3}{3}\left( \a\m+ \llm\dfrac{\c}{\alpha}\right)^3
        \\\\
        \llm =&\beta\alpha\left( \b\m+\n\dfrac{\c}{\theta}\right)-\dfrac{\beta^3\alpha^3}{3}\left( \b\m+\n\dfrac{\c}{\theta}\right)^3 
        \end{cases} .
    \end{array} 
\end{equation}

Finally, solving for $\beta$ and neglecting terms of order higher than the third, we obtain a cubic equation for $\beta$ as a function of the activation parameters $\a,\b,\c$: 
\begin{equation} 
    2\beta^3 \a\b\c+\beta^2(\a^2+\b^2+\c^2)-1=0 .
\end{equation}
The maximal noise the model can afford is the only real, positive solution of the aforementioned cubic equation that matches those reported in the phase diagrams of Fig.~\ref{fig:TAM-PD} at $\gamma = 0$.

\section{Three-directional associative memory (TAM): Applications}\label{Sec:Applications}

Once we know which are the boundaries of the TAM's retrieval region, we can confine its neural dynamics within that region and deepen its emergent computational capabilities.

\subsection{Task 1: Generalized pattern reconstruction}\label{Sec:Application-1}

\begin{figure}[tb]   
\centering
    \includegraphics[width=15cm]{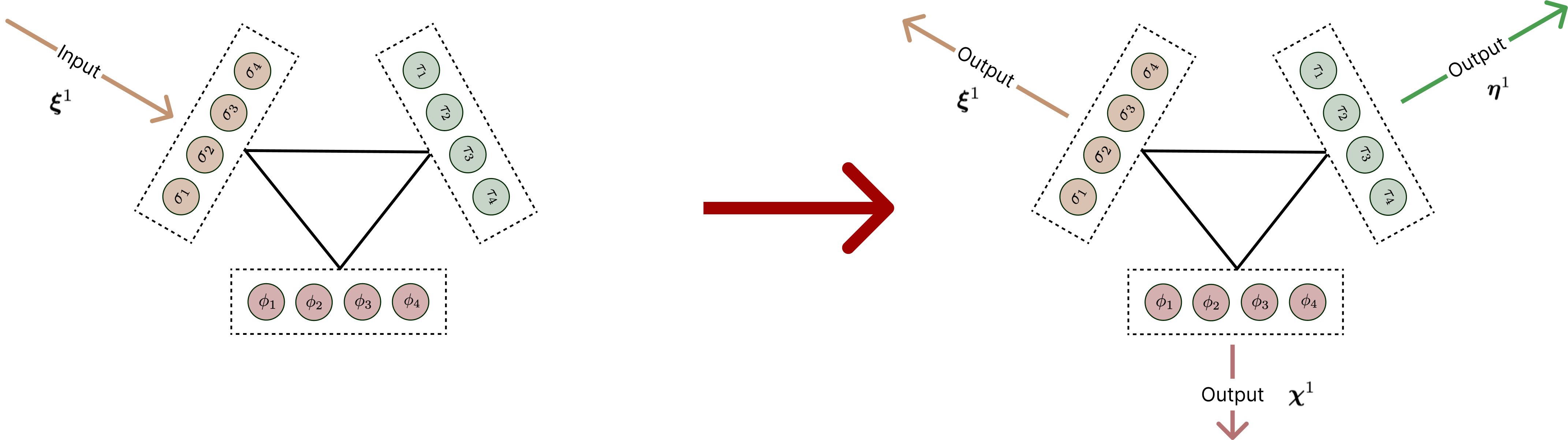}
    \caption{Schematic representation of the TAM network performing generalized pattern recognition. Left: Noisy information regarding the pattern $\bm \xi^1$ is presented on the $\sigma$-layer only. Right: As the network correctly retrieves $\xi^1$ on the $\sigma$-layer, its corresponding grandmother hidden unit (not shown here) will get active and trigger the retrieval of $\eta^1$ and $\chi^1$ on, respectively, the $\tau$-layer and the $\phi$-layer.}    
    \label{fig:TAM-ret}
\end{figure}

We start with the simplest task, that is generalized pattern reconstruction. Unlike networks equipped with solely one visible layer, here the presence of three visible layers allows for the simultaneous retrieval of three patterns, one per layer.
More precisely, by setting as initial configuration of one layer a noisy version of one of the related patterns -- say, without loss of generality, that we prepare the $\sigma$-layer  aligned with $\bm \xi^1$ apart from a fraction of flipped entries -- and setting the other two layers in a random (paramagnetic) configuration, the network can successfully reconstructs the triplet $(\bm \xi^1,\ \bm \eta^1,\ \bm \chi^1)$. In other words, we are providing as a Cauchy condition for the neural dynamics \eqref{BGmeasure} a noisy version of $\bm \xi^1$ and check whether the system spontaneously relaxes in a state where $\mathcal P(\bm \sigma, \bm \tau, \bm \phi| \bm \xi, \bm \eta, \bm \chi)$ is peaked at (or, still, very close to) the configuration $(\bm \xi^1, \bm \tau^1, \bm \chi^1)$.

\begin{figure}[tb]
    \centering
    \includegraphics[width=15cm]{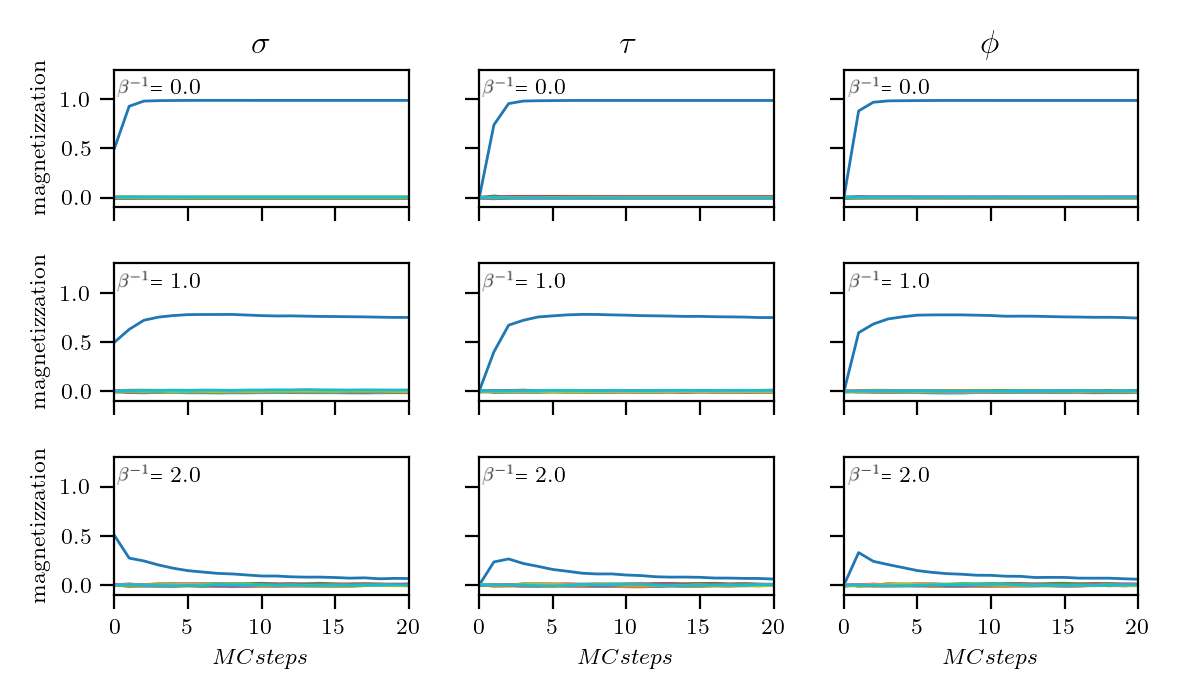}
    \caption{Monte Carlo simulations for testing TAM's ability in generalized patters recognition. In these experiments we feed the layer $\sigma$ with a noisy version of $\bm \xi^1$ (one third of the pixels are misaligned in this particular input signal), while the $\tau$ and $\phi$ layers are fed with white noise (this can be seen by the low value of their initial Mattis magnetizations). Runs are performed at $N=L=M=5000 $, $K =500$ and at different noise levels (from top to bottom: $\beta^{-1}=0, 1.0, 2.0$). Note that, as long as  $\beta^{-1}$ is small enough, beyond denoising the pattern $\bm \xi^1$ on the $\sigma$-layer, the network correctly retrieves $\bm \eta^1$ and $\bm \chi^1$ on the other two layers.}
    \label{fig:enter-sandman}
\end{figure}

\begin{figure}
    \centering
    \includegraphics[width=10cm]{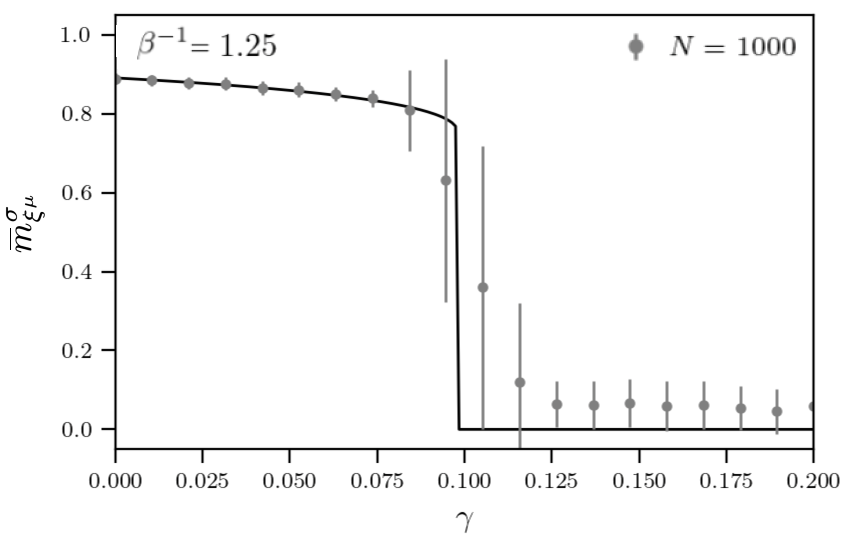}
    \caption{We compare theoretical and computational results for the Mattis magnetization $\m$. More precisely, the solid line was obtained by solving numerically the self-consistency equation \eqref{eq:self}, while the grey dots were obtained by running Monte Carlo simulations (the same plot holds also for the two other layers, therefore for $\n$ and $\llm$); these data are reported versus the storage $\gamma$. In both cases we set $\beta=0.8$ and $\alpha = \theta=1$. As for the simulations, we also need to set the network sizes ($N=M=L=1000$, with $K=\gamma N$) and the number of experiments: each point is the average over $100$ simulations, the error-bars correspond to the related standard deviation. As expected, the error-bars sensibly increase around $\gamma_c=0.10$, that is where a phase transition (bounding the retrieval region) occurs.}
    \label{fig:MC-vs-self}
\end{figure}

In Fig. \ref{fig:TAM-ret} we provide a visual representation of the task, while in Fig. \ref{fig:enter-sandman} we show the temporal evolution of the system for different choices of the parameter $\beta$ and in Fig.~\ref{fig:MC-vs-self} we present a comparison between the theoretical and the computational predictions.

\subsection{Task 2: Disentanglement of heterogeneous mixtures}\label{Sec:Application-2}
In this section,  we feed the TAM network with mixtures of patterns and check if it can return the original patterns contributing to the mixture. In particular, we shall focus on mixtures of the form $\bm \zeta \equiv \textrm{sign}(\bm \xi^1+\bm \eta^1+\bm \chi^1)$.
\newline
Before proceeding, let us point out that there is a remarkable difference in disentangling a ``heterogeneous mixture'' as $\textrm{sign}(\bm \xi^1+\bm \eta^1+\bm \chi^1)$ and disentangling a ``homogeneous mixture'' as $\textrm{sign}(\bm \xi^1+\bm \xi^2+\bm \xi^3)$\footnote{These kinds of configuration are also known as symmetric spurious states: they constitute (unwanted) stable states in the standard Hopfield model and possibly impair retrieval, see e.g., \cite{Amit,Coolen}.}. For the latter we would have a unique dataset shared by the three layers, in such a way that interference among patterns is expected to make the task more challenging. Another way to see this is by noticing that when each layer is dedicated to a different dataset, in the integral representation of the model (see Fig. \ref{fig:TAM-PD}) each visible layer is associated to two conjugated hidden layers, on the other hand, when the three visible layers share the same dataset, the two latent layers coalesce (the hidden neurons are only $3K$ instead of $6K$) and this sensibly frustrates the network. We refer to a forthcoming paper for an exhaustive discussion on disentangling homogeneous mixtures \cite{ABCRT2025}.
\newline
We also emphasize that the current task can be seen as a special case of pattern reconstruction as, for each layer, the supplied mixture $\bm \zeta$ is basically a vector correlated with one of the related patterns as, for $N, M, L \gg 1$, we can write $\bm \zeta \cdot \bm \xi^{\mu} \approx \bm \zeta \cdot \bm \eta^{\mu} \approx \bm \zeta \cdot \bm \chi^{\mu} \approx \frac{1}{2}\delta_{\mu,1}$, therefore, the input on the $\sigma$-layer can be seen as a corrupted version of the pattern $\bm \xi^1$ and similarly for the other layers.

\begin{figure}[tb]  
\centering
   \includegraphics[width=15cm]{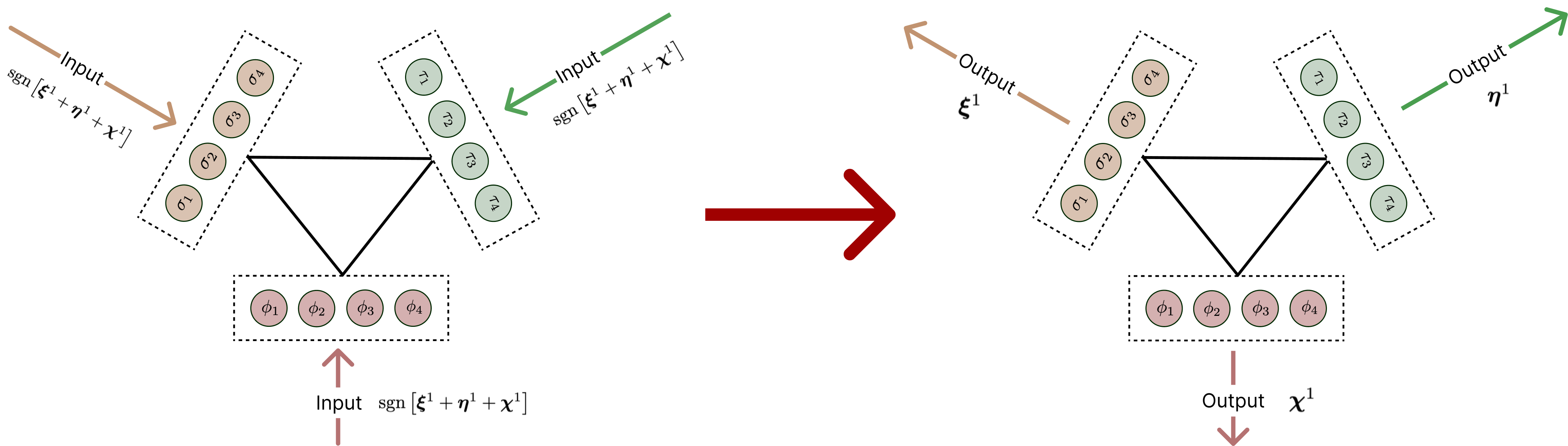}
    \caption{Schematic representation of the TAM network performing heterogeneous pattern disentanglement. Left: The mixture $\bm \zeta$ is presented to all the layers of the network. Right: All the layers recognize their corresponding patterns and the neural configurations relax to, respectively, $\bm \eta^1$, $\bm \chi^1$, and $\bm \xi^1$, so that the input mixture is successfully disentangled.}
    \label{fig:TAM-DisentaglementEye}
\end{figure}

\begin{figure}
    \centering
    \includegraphics[width=15cm]{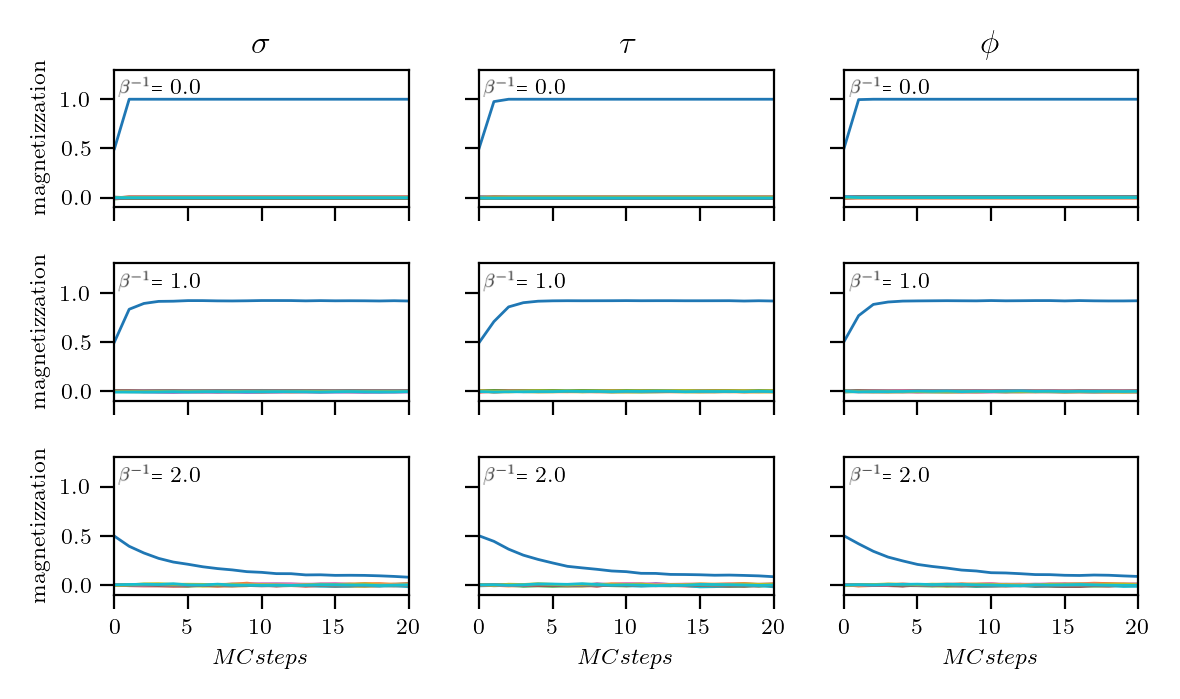}
    \caption{Monte Carlo simulations for testing TAM's ability in pattern disentanglement. Each layer is built of by $N=M=L=1000$ neurons, the number of stored pattern is $K=12$ for each dataset, and couplings are $(\a,\b,\c)=(1,1,1)$. The layers are all fed with the same spurious mixture $\bm \zeta$, which constitutes the Cauchy condition for the neural dynamics.
    Different noise levels are considered: noiseless dynamics in the first raw (where outcomes from signal-to-noise are numerically confirmed), noisy dynamics at $\beta=1$ in the second raw (where the network is still successful at the task) and noisy dynamics at $\beta=0.5$ in the third raw (where the network fails). 
    Note how, in the first two lines, the network separates the various components of the signal, projecting them on their pertinent layer such that the $\sigma$ layer reconstructs $\bm \xi^1$, the $\tau$ layer $\bm \eta^1$ and the $\phi$ layer $\bm \chi^1$. }
    \label{fig:disentanglementMC}
\end{figure}

 As in the previous section, we provide a visual representation of the task, see Fig.~\ref{fig:TAM-DisentaglementEye}, along with a plot showing the evolution of the system, where the network correctly separates the various patterns as long as the control parameters are suitably set, see Fig. \ref{fig:disentanglementMC}.

\subsection{Task 3: Disentanglement of homogeneous mixtures}\label{Sec:Penalty}

In sec.~\ref{Sec:Application-2} we disentangled mixtures of patterns, say $\bm \xi^{1}$, $\bm \eta^{1}$, $\bm \chi^{1}$, where the mixed patterns were drawn from different ``baskets'', each specific for a given layer, and we emphasized that, when the mixed patterns are sampled from the same basket,  say, $\bm \xi^{1}$, $\bm \xi^{2}$, $\bm \xi^{3}$, yielding $\bm \xi^{(1,2,3)} = \textrm{sign}(\bm \xi^1 + \bm \xi^2 + \bm \xi^3)$, 
the task is much harder and the current model is not adequate (for instance, nothing would prevent the three layers from retrieving the same pattern thus failing to accomplish this kind of task).

\begin{figure}[t]
    \centering
    \includegraphics[width=15cm]{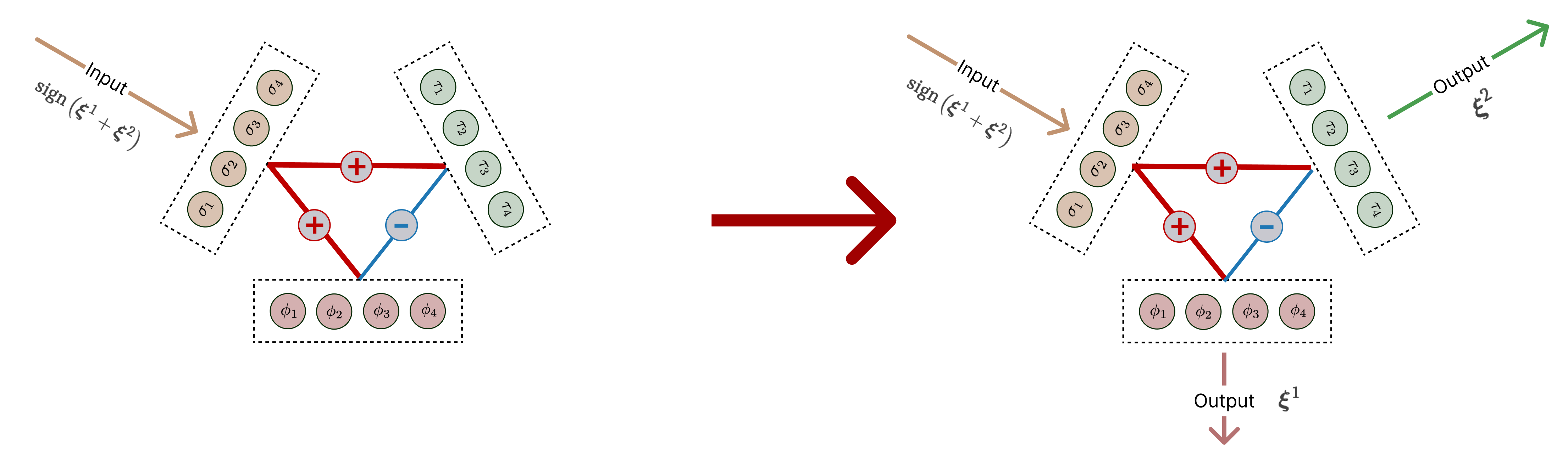}
    \caption{Schematization of the TAM used for disentangling homogeneous mixtures $\bm \xi^{(1,2)}$ made of two patterns. The layer $\sigma$ is employed to collect the input and it is clamped, the layers $\tau$ and $\phi$ return the output. The red connections highlight Hebbian-like interactions and the blue ones indicate anti-Hebbian interactions. Of course, in this scenario, we are handling a unique dataset $\bm \xi$, thus the sizes of the layers coincide ($N=M=L$).
    }
    \label{fig:TAM_with_penal}
\end{figure}

Nonetheless, as we will show, with the current model we can still tackle the disentanglement of homogeneous mixtures, provided that the number of mixed patterns is reduced from three to two.
Thus, let us consider a mixture of two patterns $\bm \xi^{(1,2)} = \text{sign}(\bm\xi^1 + \bm\xi^2)$\footnote{In the case where the argument of the sign function is zero, we assign value $+1$ or $-1$ with equal probability.} and specialize the model architecture to display one input layer and two output layers: the input is clamped as $\bm \sigma \equiv \bm \xi^{(1,2)} = \text{sign}(\bm\xi^1 + \bm\xi^2)$, and the desired outputs are $\bm \tau = \bm \xi^1$ and $\bm \phi = \bm \xi^2$. 
Moreover we need to overcome the possible scenario in which the two output layers are aligned with the same pattern: to favour the retrieval of different patterns by each output layer, we  set $\c <0$. Intuitively, this forces the output layers to work in opposition, enhancing the separation of information. As a result, the network now incorporates both Hebbian and anti-Hebbian interactions: those connecting the input and the output layers are Hebbian, while those between the output layers are anti-Hebbian. In Fig.~\ref{fig:TAM_with_penal}, we present a schematic representation of this setting. 

Referring to \eqref{H-TAM-riscritta}, the working assumption is now $\a \equiv \b \equiv g_{\sigma}>0$ while $\c \in (-\infty, 0]$ is a tunable parameter, in such a way that the model's Hamiltonian reads as: 
\begin{equation}
\label{eq:dis_hamiltonian_magn}
\begin{array}{lll}
     \mathcal H_N(\bm\tau,\bm\phi|\bm \xi) &=& - \dfrac{1}{N} \SOMMA{\mu=1}{K}\SOMMA{i,j=1}{N}\xi_i^\mu\xi_j^\mu \sigma_i^* \Big( \tau_j + \phi_j\Big) + \dfrac{|\c|}{N} \SOMMA{\mu=1}{K}\SOMMA{i,j=1}{N} \xi_i^\mu\xi_j^\mu\tau_i\phi_j
\end{array}
\end{equation}
where we set $g_{\sigma} \equiv 1$ and where the superscript ``$*$" denotes that the corresponding layer is clamped, meaning that no neural dynamics occurs in that layer. The previous Hamiltonian \eqref{eq:dis_hamiltonian_magn} can be recast using Eqs.\eqref{eq:ma}--\eqref{eq:ma_t}--\eqref{eq:ma_p} as
\begin{equation}
\label{eq:dis_hamiltonian}
\begin{array}{lll}
     \mathcal H_N(\bm\tau,\bm\phi|\bm \xi) &=& - N \SOMMA{\mu=1}{K} (m_{\xi^\mu}^\sigma)^*\Big( m_{\xi^\mu}^\tau + m_{\xi^\mu}^\phi\Big) + |\c| N \SOMMA{\mu=1}{K} m_{\xi^\mu}^\tau m_{\xi^\mu}^\phi.
\end{array}
\end{equation}

Before facing the statistical mechanics of this system, we perform some preliminary analysis to highlight the role of the parameter $\c$. In particular, we aim to determine the optimal value of $\c$ for the strength of the anti-Hebbian link, that yields perfect separation of the input spurious state. 
We will utilize a technique introduced in \cite{Pedreschi1,Pedreschi2} to study how the Mattis magnetizations of the output layers evolve after one step of neural dynamics. Given the input layer clamped to $\bm\xi^{(1,2)}$, we seek for the value of $\c$ corresponding to the target output\footnote{As the network is symmetric under the exchange $\bm\tau \to \bm\phi$, for accomplish our task, if the input layer is clamped as $\bm\sigma\equiv\bm\xi^{(1,2)}$, both $(\bm\tau, \bm\phi)\equiv (\bm\xi^1, \bm\xi^2)$ and $(\bm\tau, \bm\phi)\equiv (\bm\xi^2, \bm\xi^1)$ are optimal output states}:
\begin{equation}
    \bm m_{\xi^{\mu}}^{\tau}\sim (1, 0, \hdots, 0), \;\;\; \bm m_{\xi^{\mu}}^{\phi}\sim (0, 1, \hdots, 0)\;\;\;\mathrm{or}\;\;\;\bm m_{\xi^{\mu}}^{\tau}\sim (0, 1, \hdots, 0), \;\;\; \bm m_{\xi^{\mu}}^{\phi}\sim (1, 0, \hdots, 0).
    \label{eq:out_conf_goal}
\end{equation}
As explained in App.~\ref{app:S2N_0}, in the limit $\beta \to \infty$, the neural dynamics \eqref{dinamical} can be recast in terms of the evolution of the Mattis magnetizations, as 
\begin{equation} 
\label{eq:magn_evolv_tau_phi}
\begin{array}{lll}
     m_{\xi^\mu}^{\tau}(t+1):=\dfrac{1}{N}\SOMMA{i=1}{N}\xi_i^\mu \tau_i(t+1)=\dfrac{1}{N}\SOMMA{i=1}{N}\xi_i^\mu \tau_i(t)\mathrm{sign}\left[h^\tau_i(t|\bm\sigma\equiv\bm\xi^{(1,2)},\bm\phi) \tau_i(t)\right] ,
     \\\\
     m_{\xi^\mu}^{\phi}(t+1):=\dfrac{1}{N}\SOMMA{i=1}{N}\xi_i^\mu \phi_i(t+1)=\dfrac{1}{N}\SOMMA{i=1}{N}\xi_i^\mu \phi_i(t)\mathrm{sign}\left[h^\phi_i(t|\bm\sigma\equiv\bm\xi^{(1,2)},\bm\tau) \phi_i(t)\right],
\end{array}
\end{equation}
where $t$ is the discrete time step and 
\begin{eqnarray}
 h^\tau_i(t|\bm\sigma\equiv\bm\xi^{(1,2)},\bm\phi) &=& \dfrac{1}{N}\SOMMA{\mu=1}{K}\left(\SOMMA{j=1}{N}\xi_j^\mu\xi^{(1,2)}_j-|\c|\SOMMA{k=1}{N}\xi_k^\mu\phi_k\right)\xi_i^\mu,
 \\
 h^\phi_i(t|\bm\sigma\equiv\bm\xi^{(1,2)},\bm\tau) &=& \dfrac{1}{N}\SOMMA{\mu=1}{K}\left(\SOMMA{j=1}{N}\xi_j^\mu\xi^{(1,2)}_j-|\c|\SOMMA{k=1}{N}\xi_k^\mu\tau_k\right)\xi_i^\mu
\end{eqnarray}
are the internal fields acting on each neuron. 
For different initial configurations $( \bm\tau(0),\bm\phi(0))$ we are able to determine the magnetization
%

\begin{equation}
    \begin{array}{lll}
         m_{\bm x}^{ \bm y_1}(1)&\xrightarrow[]{N\gg 1}&\mathrm{erf}\left(\dfrac{\mu_1^{(  y_2,\bm x)}(\bm y_1)}{\sqrt{2\left(\mu_2^{(  y_2,\bm x)}(\bm y_1)-(\mu_1^{(  y_2,\bm x)}(\bm y_1))^2\right)}}\right),
    \end{array}
    \label{eq:erf_Mattis}
\end{equation}
which represents the alignment of the layer $\bm y_1 \in \{ \tau, \phi \}$ with respect to the pattern $\bm x \in \{\bm \xi^1, \bm \xi^2, \bm \xi^{\nu>2}\}$, evaluated at the first time step $t=1$, being $\bm y_2  \in \{\bm \xi^1, \bm \xi^2, \bm \xi^{\nu>2}\}$ the initial setting of the layer opposite to $\bm y_1$ (in fact, when updating the layer $\tau$ we need to specify only the state of the layer $\phi$, and vice versa, because the layer $\sigma$ is clamped and the state the initial configuration for $\tau$ is irrelevant as there are no intra-layer interactions). In particular, we consider the following three initial settings
\begin{itemize}
    \item $(\bm \tau (0), \bm\phi(0))\equiv(\bm\xi^1,\bm\xi^1)$
    \begin{equation}
        \begin{array}{lllll}
             \mu_1^{(\bm\xi^1,\bm\xi^1)}(\bm \tau)=\dfrac{1}{2}-|\c|,&&\mu_1^{(\bm\xi^1,\bm\xi^1)}(\bm\phi)=\dfrac{1}{2}-|\c|
             \\\\
             \mu_1^{(\bm\xi^1,\bm\xi^2)}(\bm\tau)=0&&\mu_1^{(\bm\xi^1,\bm\xi^2)}(\bm\phi)=0,
             \\\\
             \mu_1^{(\bm\xi^1,\bm\xi^\nu)}(\bm\tau)=0&&\mu_1^{(\bm\xi^1,\bm\xi^\nu)}(\bm\phi)=0,
        \end{array}
        \label{eq:111_1}
    \end{equation}
    \item $( \bm\tau(0),\bm\phi(0))\equiv(\bm\xi^1,\bm\xi^2)$
    \begin{equation}
        \begin{array}{lllll}
             \mu_1^{(\bm\xi^2,\bm\xi^1)}(\bm\tau)=\dfrac{1}{2},&&\mu_1^{(\bm\xi^1,\bm\xi^1)}(\bm\phi)=\dfrac{1}{2},
             \\\\
             \mu_1^{(\bm\xi^2,\bm\xi^2)}(\bm\tau)=0,&&\mu_1^{(\bm\xi^1,\bm\xi^2)}(\bm\phi)=0,
             \\\\
             \mu_1^{(\bm\xi^2,\bm\xi^\nu)}(\bm\tau)=0,&&\mu_1^{(\bm\xi^1,\bm\xi^\nu)}(\bm\phi)=0,
        \end{array}
        \label{eq:112_1}
    \end{equation}   
    \item $(\bm\tau(0),\bm\phi(0))\equiv(\bm\xi^1,\bm\xi^{\nu>2})$
    \begin{equation}
        \begin{array}{lllll}
             \mu_1^{(\bm\xi^\nu,\bm\xi^1)}(\bm\tau)=\dfrac{1}{2},&&\mu_1^{(\bm\xi^1,\bm\xi^1)}(\bm\phi)=0,
             \\\\
             \mu_1^{(\bm\xi^\nu,\bm\xi^2)}(\bm\tau)=0,&&\mu_1^{(\bm\xi^1,\bm\xi^2)}(\bm\phi)=0,
             \\\\
             \mu_1^{(\bm\xi^\nu,\bm\xi^\nu)}(\bm\tau)=0,&&\mu_1^{(\bm\xi^1,\bm\xi^\nu)}(\bm\phi)=0,
        \end{array}
        \label{eq:113_1}
    \end{equation}
\end{itemize}
These results state that the configuration $(\bm \tau (0), \bm\phi(0))\equiv(\bm\xi^1,\bm\xi^{\nu>2})$ is not stable, while the target configuration $(\bm \tau (0), \bm\phi(0))\equiv(\bm\xi^1,\bm\xi^{2})$ is stable, however, also the configuration $(\bm \tau (0), \bm\phi(0))\equiv(\bm\xi^1,\bm\xi^{1})$
is potentially stable, but setting  $\c^* = - \frac{1}{2}$ that configuration is made as well unstable. In other words, we potentially have (at least) two attractors corresponding to the (target) disentangled configuration and to the (unwanted) configuration where the same pattern is retrieved by both layers; for the former the stability is independent of $\c$, while for the latter the stability can be destroyed by setting $\c = 1/2$.
The convenience in setting $|\c^*| = \frac{1}{2}$ is deepened in the next subsec.~\ref{bias}.





\begin{figure}
    \centering
    \includegraphics[width=10cm]{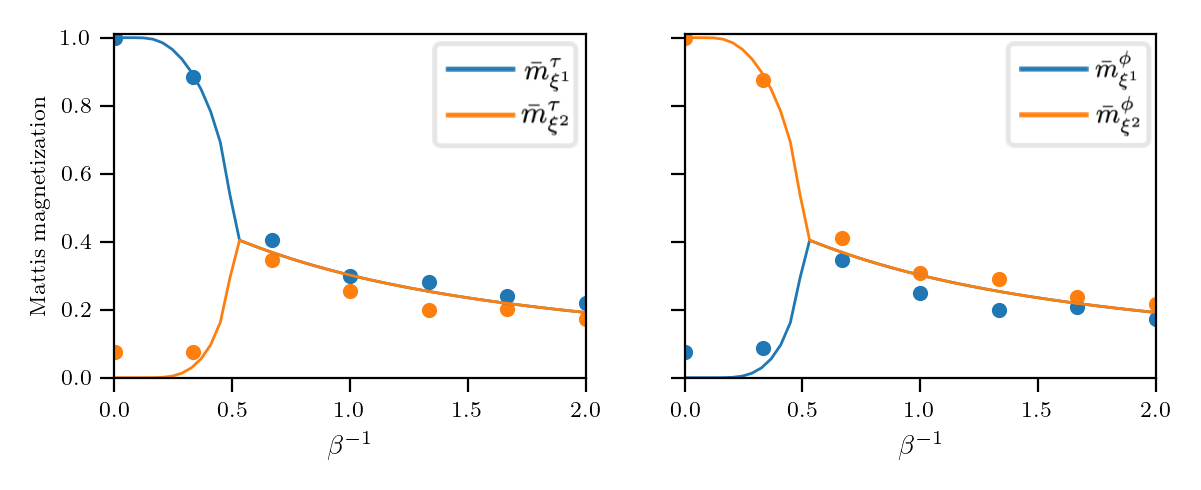}
    \caption{Dependence on $\beta^{-1}$ of the magnetizations $\bar m_{\xi^1}^{\tau}$ and $\bar {m}_{\xi^2}^{\tau}$ related to the output layer $\tau$ (left) and of the magnetizations $\bar m_{\xi^1}^{\phi}$ and $\bar{m}_{\phi^2}^{\tau}$ related to the other output layer $\phi$ (right), when the input is supplied on the layer $\sigma$ that is clamped as $\boldsymbol \sigma \equiv \bm \xi^{(1,2)}$. Monte Carlo simulations (bullets) are run setting $K=12$ and $N=M=L=1500$, that is also the length of all the patterns, and compared with results stemming from the self-consistent equations (solid lines), see \eqref{eq:scsc}.
    The coupling between output layers is $|\c|=|\c^*|=0.5$.
    Note the existence of a phase transition at $\beta^{-1} \sim 0.5$, below which the network separates the two patterns.  }
    \label{fig:phase_spur_basso}
\end{figure}

Here, to corroborate the previous results, obtained by simply statistical arguments, we now perform a statistical-mechanical analysis where the results presented in Sec.~\ref{Sec:Theory} are adopted to the current case, more precisely, the input layer is clamped as $\bm \sigma \equiv \bm\xi^{(1,2)}$ and we restrict to the low-storage regime, namely $\lim_{N\to+\infty}K/N=0$.
Thus, starting from \eqref{eq:dis_hamiltonian} by applying Guerra interpolation, under the constraint of low-load regime, we get the following self-consistency equations 
\begin{equation} \label{eq:scsc}
\begin{array}{lll}
     \bar{m}^\tau_{\xi^a}=  \mathbb{E}\left[\tanh\left(\beta\SOMMA{\nu=1}{2}\xi^{\nu}\left( (m_{\xi^\nu}^\sigma)^* -|\c|\bar{m}^\phi_{\xi^\nu}\right)\right)\xi^{a}\right]
     \\\\
     \bar{m}^\phi_{\xi^a}=  \mathbb{E}\left[\tanh\left(\beta\SOMMA{\nu=1}{2}\xi^{\nu}\left( (m_{\xi^\nu}^\sigma)^* -|\c|\bar{m}^\tau_{\xi^\nu}\right)\right)\xi^{a}\right],
\end{array}
\end{equation}
 where we use the notation $(m_{\xi^\nu}^\sigma)^*$ to indicate the Mattis magnetization of the clamped layer $\sigma$ with respect to the pattern $\bm\xi^\nu$.  
The solutions are shown in the plots of Fig.~\ref{fig:phase_spur_basso}, together with Monte Carlo runs, that perfectly fit the analytical findings.

\subsubsection{Biased dataset} \label{bias}
\begin{figure}[t]
    \centering
    \includegraphics[width=12cm]{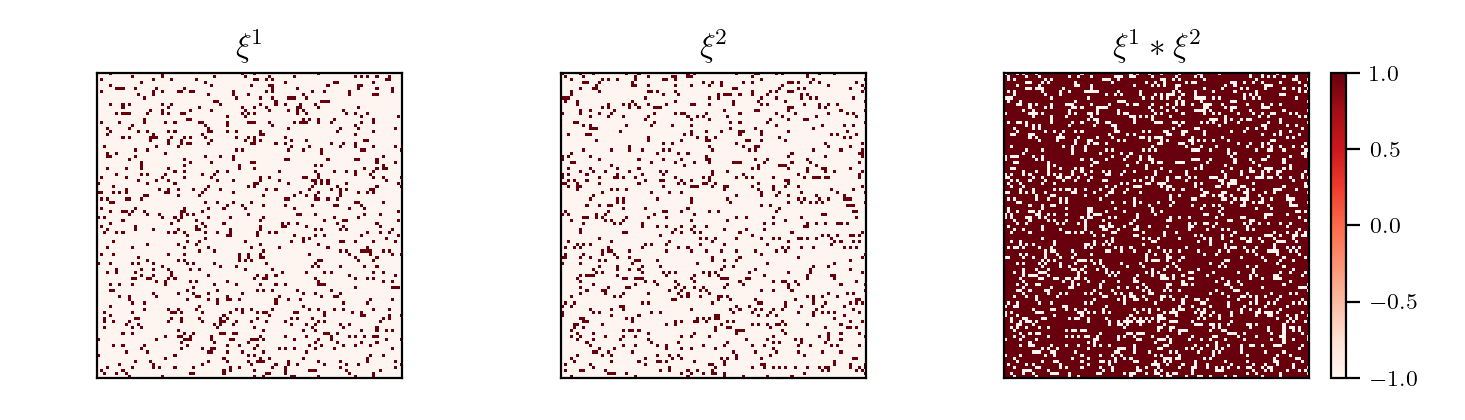}
    \caption{Example of biased patterns, with $b=0.8$. In the first two panels we report two patterns --$\xi^1$ and $\xi^2$-- that, being largely composed of $-1$ values, appear as almost blank images and in the last panel we show their product $\xi^1 \cdot \xi^2$ that, being composed of mainly $+1$ values, appears as an almost black picture.}
    \label{fig:example_bias_patt}
\end{figure}

Analogous results can be obtained in the case where patterns display a bias making one of the two entries more likely, see Fig.~\ref{fig:example_bias_patt} for an example of biased dataset.
Interestingly, in this case one finds that the optimal choice of $\c$ is again $0.5$ and thereby $\c^*$ is load and dilution independent.

Let us model a biased dataset by perturbing the standard Rademacher distribution of the entries of our patterns, introducing a parameter $b \in [0,1]$:
\begin{equation}
    \mathbb{P}(\xi_i^\mu)=\dfrac{1+b}{2}\delta(\xi_i^\mu + 1)+\dfrac{1-b}{2}\delta(\xi_i^\mu - 1)\;\;\;\;
\end{equation}
for $i=1,\hdots, N$ and  $\mu = 1, \hdots,K$.
This prescription ensures that, given two patterns $\bm{\xi}^\mu$ and $\bm{\xi}^\nu$, the parameter $b$ determines their average activity level and correlation:
\begin{equation}
    \begin{array}{lll}
         \mathbb{E}\left[\xi_i^\mu\right]=-b, && \mathbb{E}\left[\xi_i^\mu\xi_j^\nu\right]=\delta_{i,j}\left[\delta_{\mu\nu}+b^2(1-\delta_{\mu\nu})\right].
         \label{eq:bias_rules}
    \end{array}
\end{equation}
Then, we simply adapt the AGS low-firing coding level technique \cite{AGS-87,ALM-21} to the case, namely we revise the Hamiltonian \eqref{eq:dis_hamiltonian} as
\begin{equation}
\begin{array}{lll}
    \mathcal H_{N,b}(\bm\tau,\bm\phi|\bm \xi) &=& - N \SOMMA{\mu=1}{K} (m_{\xi^\mu}^\sigma)^*\Big( m_{\xi^\mu}^\tau + m_{\xi^\mu}^\phi\Big) + |\c| N \SOMMA{\mu=1}{K} m_{\xi^\mu}^\tau m_{\xi^\mu}^\phi
    \\\\
     &&+ g N \left[\left(M^{\bm\tau}+b\right)^2+\left(M^{\bm\phi}+b\right)^2\right] 
     \label{eq:Hamil_bias}
\end{array}
\end{equation}
where we posed
$ M^{\bm y} = \dfrac{1}{N}\SOMMA{i=1}{N}  y_i $ and $\tilde{m}_{\xi^\mu}^{\bm y} = \dfrac{1}{N}\SOMMA{i=1}{N} \left(\xi_i^\mu +b\right) y_i = m_{\xi^\mu}^{\bm y} + b M^{\bm y}$.
The third term on the r.h.s. of Eq. \eqref{eq:Hamil_bias} is needed to force the average magnetization of each family of spins to be equal to $-b$.

\begin{figure}[t]
    \centering
    \includegraphics[width=12cm]{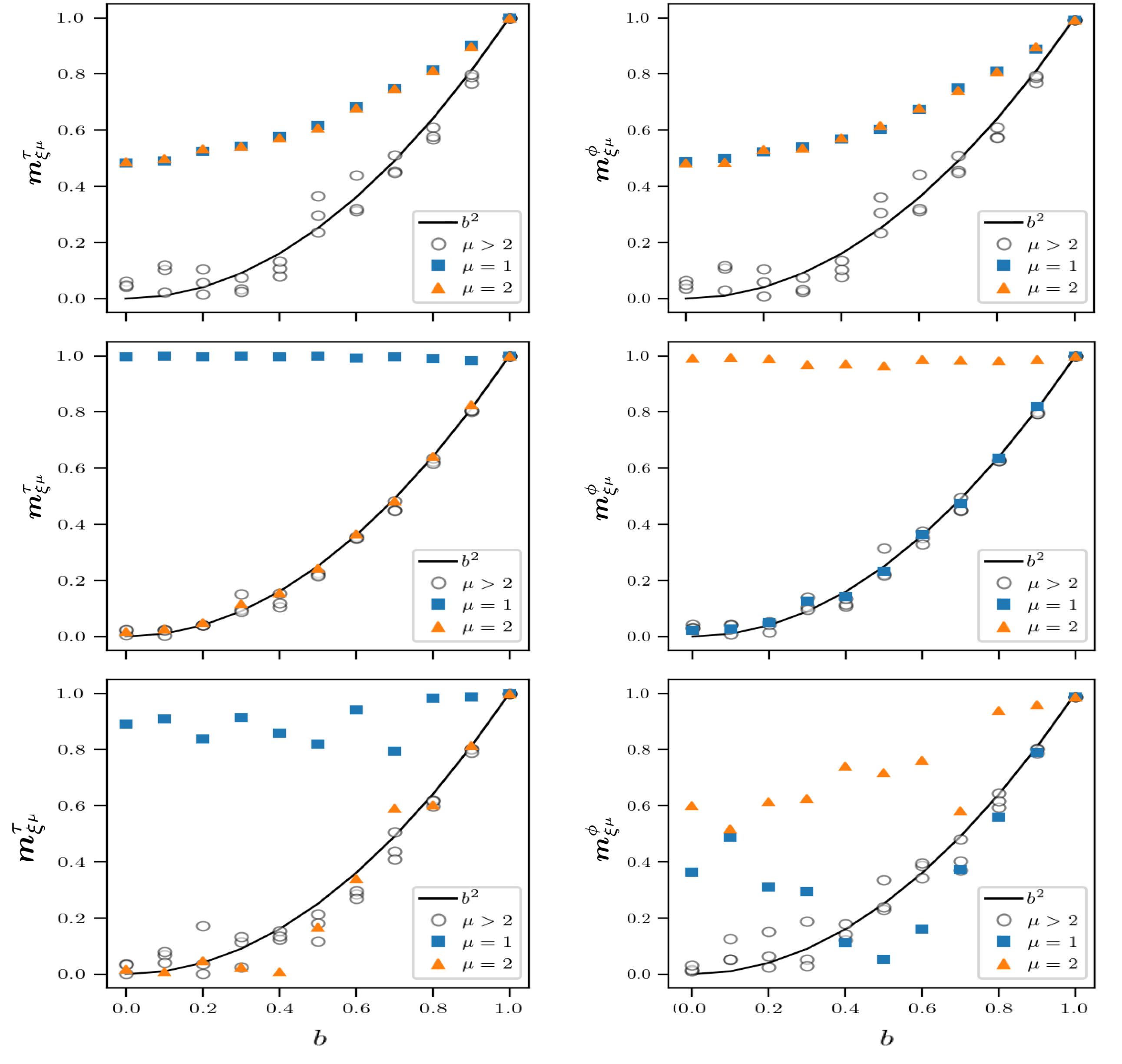}
    \caption{Monte Carlo simulations for testing TAM's ability in pattern disentanglement in the case of biased dataset. Different columns refer to different output layers ($\tau$ on the left and $\phi$ on the right), while the layer $\sigma$ is clamped as $\mathrm{sign}(\bm\xi^{1}+\bm\xi^{2})$ and used as input layer. Different rows correspond to different values of the interaction strength $\c$: a soft separation constant $|\c|=0.1$ (upper row), the optimal separation constant $|\c^*|=0.5$ (middle row) and a too strong separation constant $|\c|=1.5$ (lower row). 
In each panel we report the Mattis magnetization for the related layer evaluated with respect to $\bm \xi^1$ ($\square$), to $\bm \xi^2$ ($\triangle$) and to any $\bm \xi^{\nu>2}$ ($\bigcirc$), plotted versus the bias parameter $b=0$; the case $b=0$ corresponds to no bias. In the middle row, the proper Mattis magnetizations are stable on values $\approx 1$ for all the $b$ values reported in the abscissa, furthermore the Mattis magnetization related to non retrieved patterns show a parabolic shape in $b$  as expected given the correlation among patterns. Note also that too strong or too small values of $|\c|$ impair the retrieval (upper and lower panels respectively). In all the simulations we set $N=500$ and $K=10$.}
\label{fig:termalizzazione_lambda_star}
\end{figure}

\begin{figure}[t]
    \centering
    \includegraphics[width=12cm]{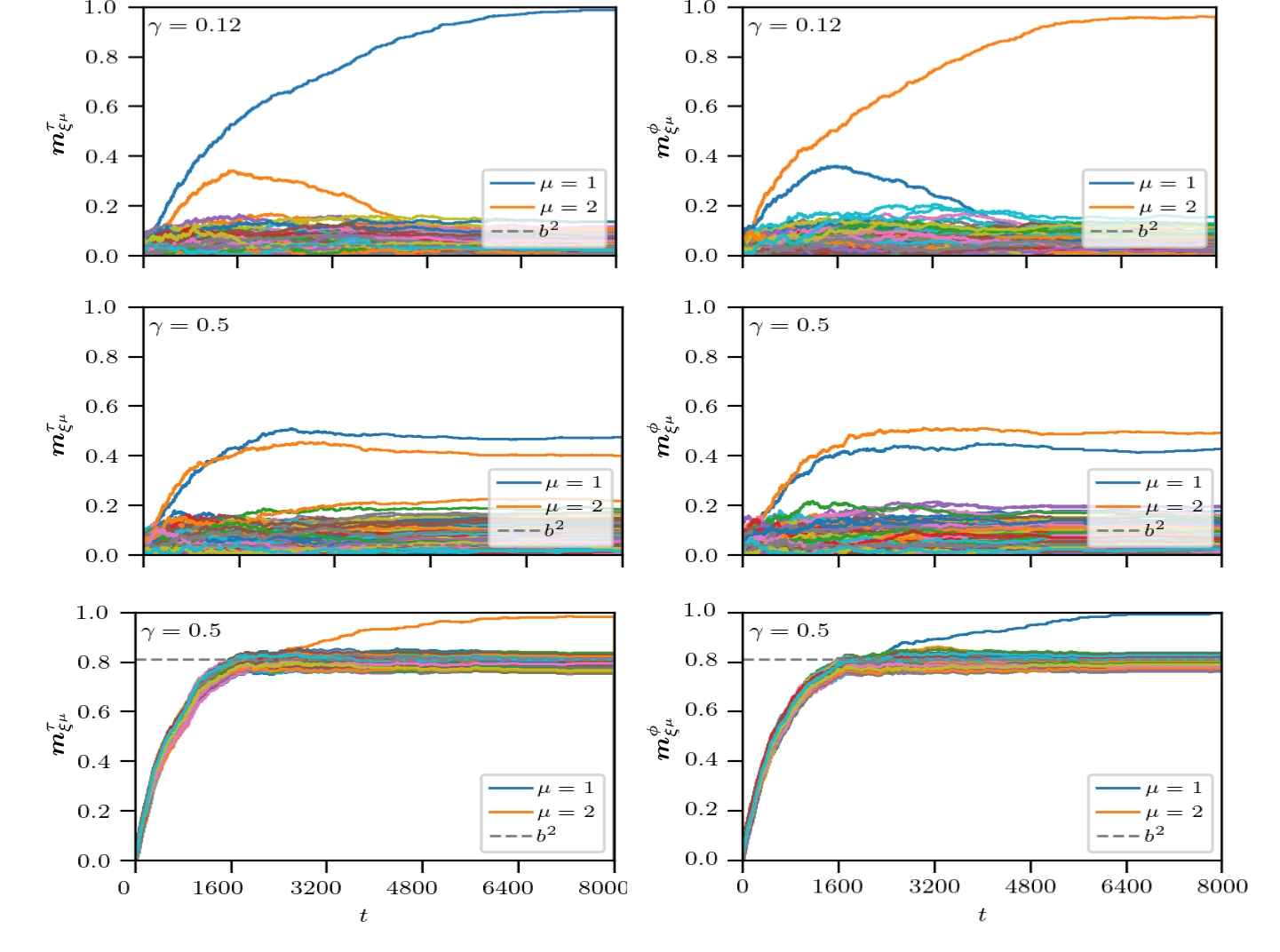}  
    \caption{Temporal evolution of the Mattis magnetizations $m_{\xi^\mu}^{\tau, \phi}$ for the output layers $\tau$ (right) and $\phi$ (left) during a noiseless dynamics ($\beta^{-1}=0$) with the input layer $\sigma$ clamped to $\mathrm{sign}(\bm\xi^1+\bm\xi^2)$ and $|\c|=|\c^*|=0.5$.
    Lower panels: $(\gamma,b)=(0.5,0.85)$  i.e., high load and high bias. The TAM raises all the magnetizations (as they are all extremely correlated among each other because $M=0.85$), yet, after approximately $2 \cdot 10^3$ Monte Carlo steps, the correct magnetizations raise from $\sim 0.85$ to $\sim 1$.
    Middle panels: $(\gamma,b)=(0.5,0.2)$ i.e. high load and low bias. The two Mattis magnetizations are simultaneously stuck at values $\sim 0.5$, thus none of the mixed patterns is retrieved.  This scenario is more challenging than the previous one, where the extremely high values of bias (i.e., $b = 0.85$) made the content of information per pattern minimal, conversely, here relatively low values of bias (i.e., $b=0.2$) makes the dataset close to the Rademacher reference, hence the retrieval region is expected to resemble that of the random case provided in the phase diagrams of Fig.~\ref{fig:TAM-PD} with $\gamma_c \approx 0.33$.
    Upper panels: $(\gamma,b)=(0.12,0.2)$ i.e. low load and low bias. This is a scenario affordable by the network. Indeed, after a transient where in both layers the two Mattis magnetizations (related to the two patterns $\bm\xi^1$ and $\bm\xi^2$) are raised simultaneously (due to their correlated inner structure), a separation occurs at approximately $2 \cdot 10^3$ Monte Carlo steps, such that the correct signal is retrieved on each layer. For all these simulations we retained $N=M=L=1500$ and $K=\gamma N$.}
    \label{fig:separation_carico_bias}
\end{figure}

Detailed calculations leading to $|\c^*| = 1/2$ are presented in App.~\ref{sec:S2N_biased}, while here we corroborate this result by Monte Carlo simulations, see Fig.s \ref{fig:termalizzazione_lambda_star} and \ref{fig:separation_carico_bias}.

\subsection{Task 4: Retrieval of temporal sequences of patterns}\label{Sec:Application-3}
\begin{figure}[t]
    \centering
    \includegraphics[width=8cm]{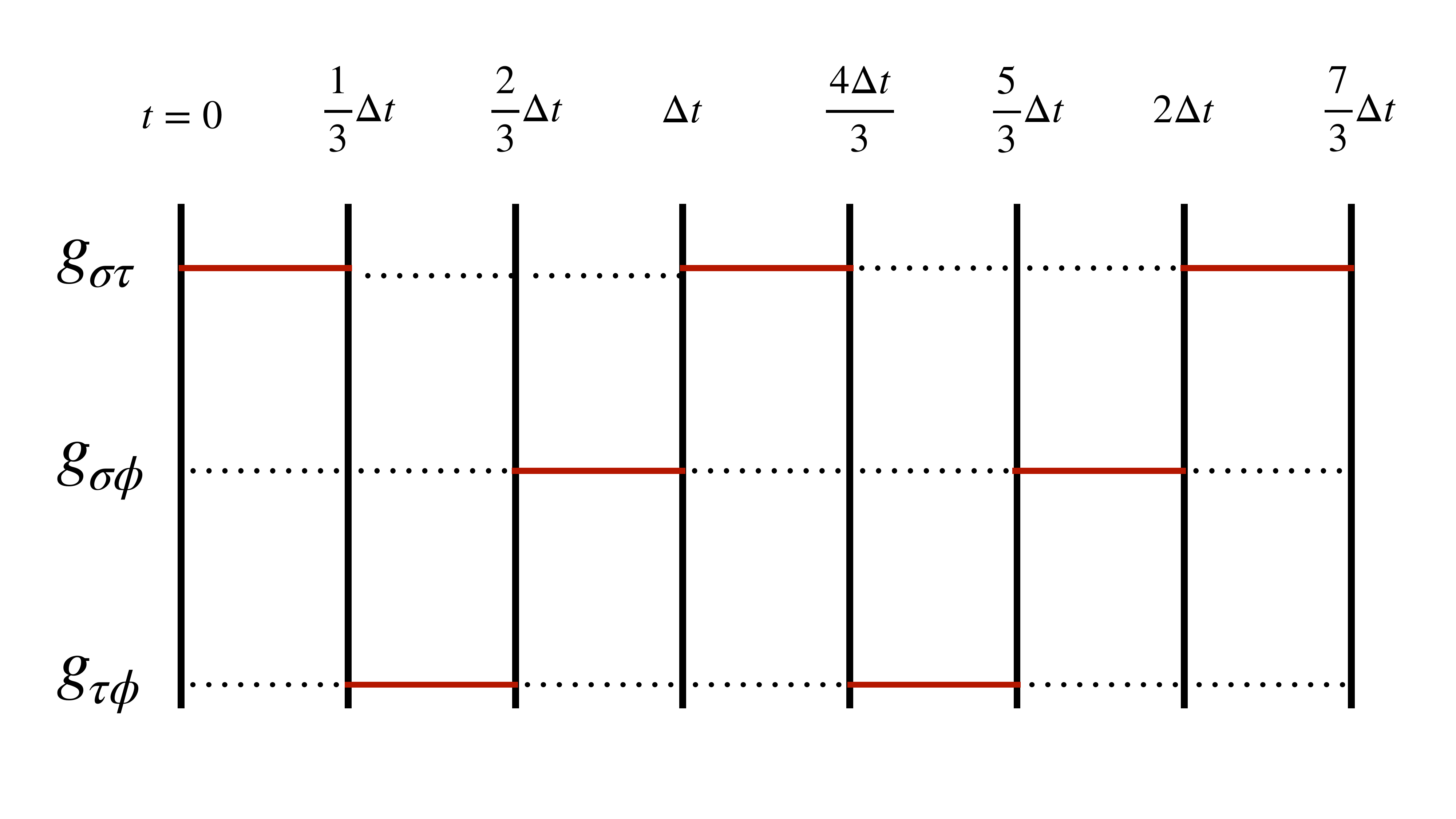}
    \caption{Schematic representation of the activation and deactivation of inter-layer connections. At any instant of time only one pair of layers is connected, in such a way that one layer remains detached. A given inter-layer connection, say $\a$ remains active for a time span equal to $\Delta t$, then it is set quiescent and $\c$ is activated. Again, the latter remains active for a time span equal to $\Delta t$, then it is set quiescent and $\b$ is activated, and so on so forth. }
    \label{fig:BAM_cycles}
\end{figure}

\begin{figure}[t]
    \centering
    \includegraphics[width=15cm]{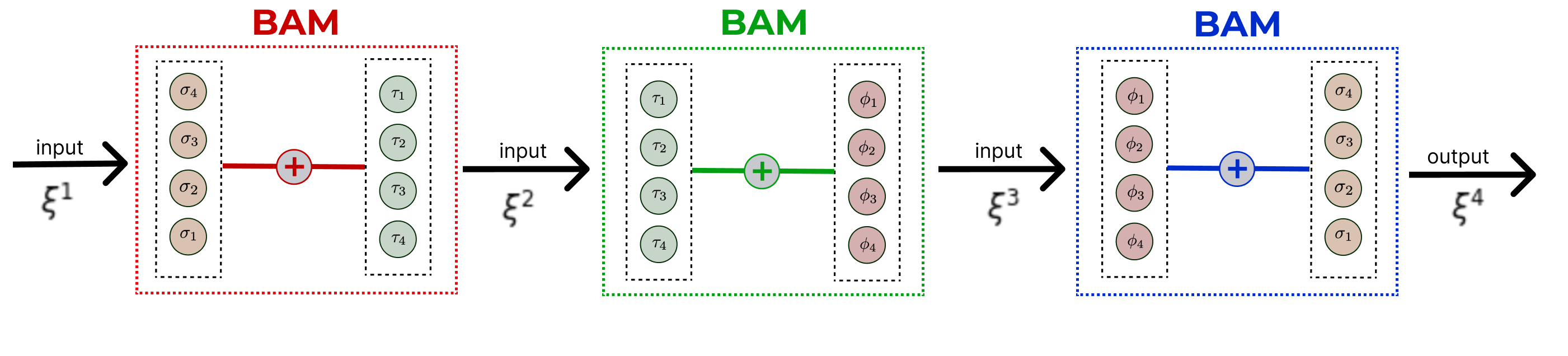}
      \includegraphics[width=15cm]{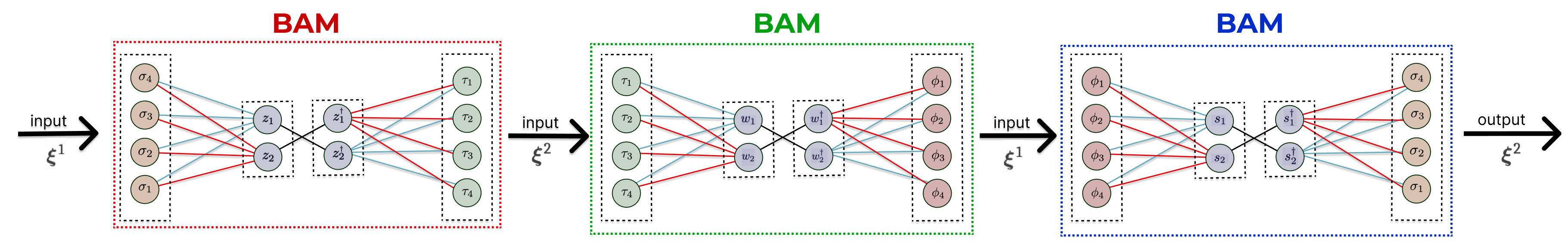}
    \caption{Top: Sketch of the TAM's architecture stemming from the rule \ref{eq:gst}, which makes layers interact in pairs, leaving one layer out by rotation. Bottom: The equivalent representation of this TAM is made of a series of coupled RBMs where, unlike the RBM obtained for a standard BAM, the hidden neuron $z_{\mu}$ is connected with the conjugate hidden neuron $z_{\mu+1}^\dag$ and periodic boundary conditions apply ($z_{K+1}^\dag = z_{1}^\dag $).}
    \label{fig:BAM_temporal1}
\end{figure}

In this section we aim to mirror extensions implemented in auto-associative networks, where patterns are retrieved consecutively according to a prescribed sequence (see e.g., \cite{AmitChimes,MezardChimes,KanterChimes}). 
Without loss of generality, let 
\begin{equation} \label{eq:sequence}
\bm \xi^1, \bm \xi^2, \bm \xi^3, ..., \bm \xi^K, \bm \xi^1, ... 
\end{equation}
the specific, periodic sequence of patterns that we want to retrieve. As in the previous sec.~\ref{Sec:Penalty}, here any layer will be employed for the same kind of patterns, that is, we have a unique dataset $\bm \xi$ in such a way that the three layers must necessarily display the same size, i.e., $N=M=L$. Then, we want the neuronal configurations of the three layers to evolve in such a way that we can read the sequence above.
The core idea is to think at the two patterns appearing in the generalized Hebbian kernels, e.g. $\bm \xi^{\mu}, \bm \eta^{\mu}$ in $\boldsymbol J^{\sigma \tau}$, as two consecutive patterns in the sequence \ref{eq:sequence} (e.g. $\bm\eta^{\mu}\equiv \bm\xi^{\mu+1}$), such that the retrieval of the pattern $\mu$ by one layer of the network triggers the retrieval of the pattern $\mu+1$ by the adjacent layer and so on.
Further, the strengths of the inter-layer interactions will be taken functions of time, in such a way that only two layers at a time are allowed to interact or, otherwise stated, one layer, by rotation, is detached. 
\newline
Let us denote by $\Delta t$ the time lapse (measured in neuronal updatings, see eq.~\ref{dinamical}) along which we want to retain the retrieval on one layer, then 
\begin{equation}\label{eq:gst}
\a(t) = \begin{cases}
    1, ~~ \textrm{if}~~  n\leq t/\Delta t \leq n+1/3\\
    0, ~~ \textrm{otherwise}~~
\end{cases}
\end{equation}
for any $n\in \mathbb N$. The periodic $\b(t),\c(t)$ factors are obtained from \eqref{eq:gst} by adding a phase of $1/3$ to the $t/\Delta t$ condition for the weight $\b$ and of $2/3$ for the weight $\c$. The rule is sketched in Fig. \ref{fig:BAM_cycles} for clarity and, as shown in Fig.~\ref{fig:BAM_temporal1}, the resulting architecture can be viewed as a sequence of BAM modules that couple nearby layers in a periodic fashion.  

\begin{figure}[t]
    \centering
    \includegraphics[width=12cm]{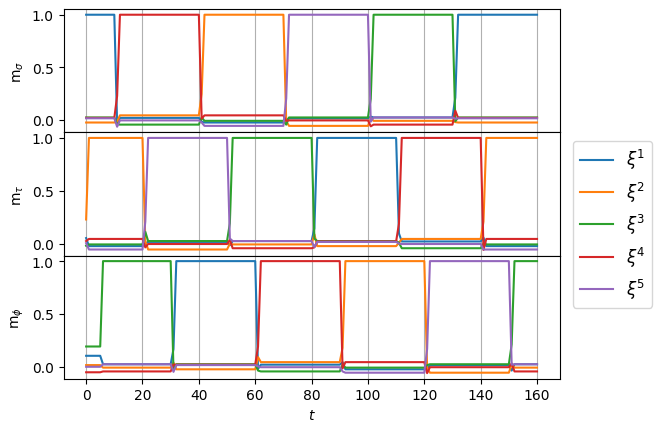}
    \caption{Retrieval of ordered sequences of alternating patterns by the TAM. The simulated network has $N=M=L=1000$ neurons per layer and it handles  $K=5$ patterns $\bm\xi^1,..,\bm\xi^5$ recursively (that is, with the periodic boundary condition $\bm\xi^6 \equiv \bm\xi^1$), further, we set $\Delta t=30$. The neural dynamics has been simulated at noise level $\beta^{-1}=0.1$. Note that, initially, the $\sigma$-layer retrieves $\bm\xi^1$ and this forces the $\tau$-layer to retrieve $\bm\xi^2$. This, in turn, triggers the retrieval of $\bm\xi^3$ by the layer $\phi$ that forces the layer $\sigma$ to retrieve $\bm\xi^4$ and so on.}
    \label{fig:sequence-TAM1}
\end{figure}

The TAM cost function for this setting reads as 
\begin{align}
   \mathcal{H}_{N}(\bm\sigma,\bm\tau,\bm\phi ; t|\bm\xi) = -\frac{1}{N} \sum_{\mu=1}^K\sum_{i,j=1}^N \lr{\a(t) \xi_i^\mu \xi_j^{\mu+1} \sigma_i \tau_j + \b(t) \xi_i^\mu \xi_j^{\mu+1} \phi_i \sigma_j + \c(t)  \xi_i^\mu \xi_j^{\mu+1}\tau_i \phi_j}.
    \label{eq:CyclesTAM-H}
\end{align}
The coefficients $\a, \b , \c$ are now used to trigger the transition from one retrieval state to another by activating and deactivating the relevant pairwise couplings. To understand how this may work, consider for instance the setting $(\a =1, \b=0, \c=0)$, which activates the first coupling in the cost function \eqref{eq:CyclesTAM-H}  and deactivates the others. At leading order in $N$, the conditional Boltzmann-Gibbs probability for $\bm \tau$, being $\bm \sigma=\bm \xi^{\mu}$, see \eqref{BGmeasure}, reads
\begin{align}
    \mathcal P(\tau_i|\bm\sigma= \bm\xi^\mu) = s\lr{2\beta\xi_i^{\mu+1} \tau_i},
\end{align}
where $s(x)=\frac{1}{1+e^{-x}}$ is the sigmoid function, and it is maximized when $\bm\tau=\bm\xi^{\mu+1}$. In this way the ``transition'' $\bm\xi^\mu \to \bm\xi^{\mu+1}$ is triggered: notice that the transition is applied on different layers, namely the layer $\bm\tau$ gets aligned with the vector $\bm \xi^{\mu+1}$, while the layer $\bm \sigma$ holds the previous configuration $\bm\sigma=\bm\xi^\mu$. By properly tuning $\a, \b, \c$ this mechanism is iterated: after a fixed number of time units $\frac{\Delta t}{3}$, the setting switches to $(\a=0,\b=0,\c=1)$, which triggers the transition $\bm \xi^{\mu+1} \to \bm \xi^{\mu+2}$, that is, the layer $\bm \phi$ gets aligned with $\bm \xi^{\mu+2}$. At this point, the system is in the ordered state $(\bm\sigma,\bm\tau,\bm\phi)=(\bm\xi^1,\bm\xi^2,\bm\xi^3)$, where it stands for a time lapse equal to $\frac{\Delta t}{3}$, after which the cycles restart by switching to the setting $(\a=0,\b=1,\c=0)$, which now forces the transition of the layer $\bm\sigma$ based on the state of the $\bm \phi$ layer, namely, $\bm \sigma=\bm\xi^{\mu+3}$. 
This procedure can be repeated until convergence to a final pattern or indefinitely by imposing periodic condition $\bm\xi^{\mu+K} = \bm\xi^{\mu}$, eventually covering the whole space of triplets $(\bm\sigma,\bm\tau,\bm\phi)=(\bm\xi^\mu,\bm\xi^{\mu+1},\bm\xi^{\mu+2})$ in a ordered temporal sequence, see Fig.~\ref{fig:sequence-TAM1}. 
\newline
We close this section with a remark. 
In the past, Amit proposed the realization of an auto-associative neural network able to count temporal cycles by perturbing the Hamiltonian of the Hopfield model with an asymmetric term that, when activated, triggers the transition $\bm\xi^\mu\to \bm\xi^{\mu+1}$ \cite{AmitChimes}; this term looks very similar to our Hamiltonian pairwise couplings, that is $\sum_{\mu}\sum_{ij}\xi_i^{\mu}\xi_j^{\mu+1}\sigma_i\sigma_j$. However, the consequences of such a contribution in the two cases are qualitatively different: in the auto-associative setting, this breaks detailed balance, while, in the hetero-associative setting, detailed balance is preserved \cite{Kosko}.
To see this, note that the TAM Hamiltonian is the sum of three equivalent terms that couple the layers in a pairwise fashion: it suffices to consider only one of them, then the same argument applies to the others terms without loss of generality.  Consider for instance the term $\mathcal{H}_{\sigma \tau} = \sum_\mu \sum_{ij}\xi_i^\mu\xi_j^{\mu+1}\sigma_i \tau_j$: it is a quadratic form in the coupling matrix $J_{ij}=\sum_\mu \xi_i^\mu\xi_j^{\mu+1}$, \emph{i.e.} $\mathcal{H}_{\sigma \tau} = \boldsymbol{\sigma}^T \boldsymbol{J} \boldsymbol{\tau}$, hence it is trivial to show that $\mathcal{H}_{\sigma \tau}^T = \boldsymbol{\tau}^T \boldsymbol{J}^T \boldsymbol{\sigma} = \sum_\mu \sum_{ji}\xi_j^{\mu+1} \xi_i^\mu \tau_j \sigma_i = \mathcal{H}_{\sigma \tau}$. On the other hand, in the case of $\bm\tau = \bm\sigma$ this would only be true if and only if $\boldsymbol{J}^T=\boldsymbol{J}$. Hence, another reward of hetero-associative neural networks is their robustness to deal with dynamical patterns.  

\begin{figure}[t]
    \centering
    \includegraphics[width=12cm]{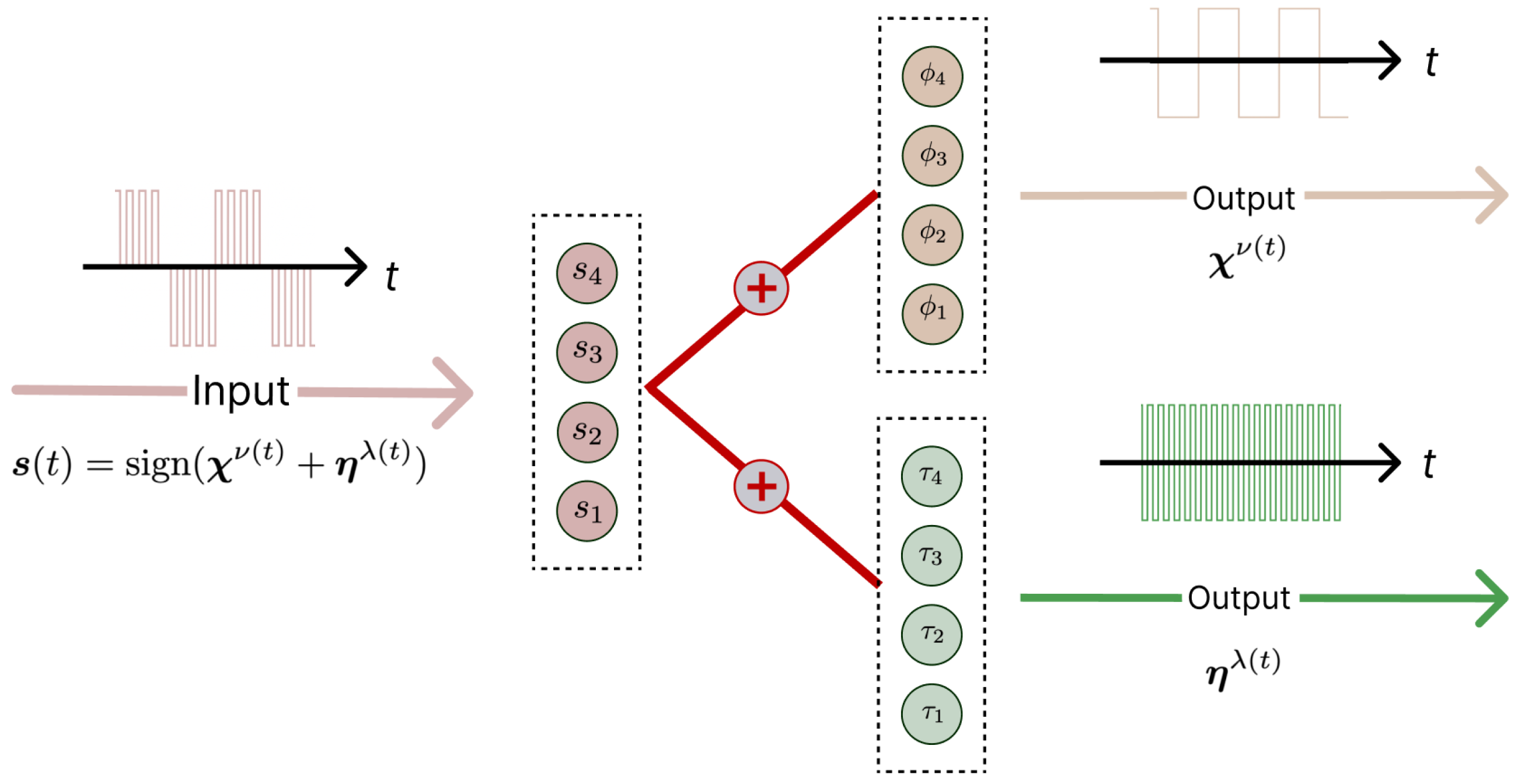}
    \caption{Sketch of the TAM's architecture stemming from the rule \eqref{FM-TAM}. The layer $s(t)$ is used to collect a time-dependent square weave as input, the layers $\phi$ and $\tau$ return as output two square waves, respectively the  ``slow'' and ``fast'' varying components of the input square-weave.}
    \label{fig:f_mod_schema}
\end{figure}

\subsection{Task 5:  Frequency modulation with mixtures of periodic patterns}\label{Sec:Application-4}
In this section we employ the TAM to decode mixtures of patterns characterized by different frequencies, a task reminiscent of {\em frequency modulation} in radio transmissions: just like radio signals are modulated according to the composition of two components, the slow (base-band) frequency and the fast (information-carrier) component, here we propose a similar mechanism operated by the TAM. 
\newline
The input is a mixture of two patterns, but, unlike the application described in the previous subsec.~\ref{Sec:Penalty}, here the patterns involved belong to different datasets and change with time at a different pace. 
Let us point out that, with this setting, the tasks the TAM has to face are, at first, the disentanglement of two periodic sequences of patterns mixed together, then the pattern recognition within each sequence such that one of the output layer returns the time-ordered sequence of the slowly evolving pattern and the other the time-ordered sequence of the fast ones. 
\newline
We inspect two prototypical scenarios, namely when the signal that we ask the network to separate has the form of a linear combination of sequences 
\begin{equation}\label{eq:s1}
    \bm s(t)=\frac{1}{2}\lr{\bm\chi^{\nu(t)}+\bm\eta^{\lambda(t)}},
\end{equation}
or when the signal is the sign of this combination (note that the sign function is non-linear):
\begin{equation}\label{eq:s2}
    \bm s(t)=\sign(\bm\chi^{\nu(t)}+\bm\eta^{\lambda(t)}).
 \end{equation}   
Here the time-dependent indices $\nu(t),\lambda(t)$ have been introduced to account for the ``slow'' and ``fast'' varying components; in particular, there are $K_1 = \gamma_1 N$ slow patterns and $K_2 = \gamma_2 N$ fast patterns and they are of the form $\nu(t)=1+t\Mod{T_1},\lambda(t)=1+t\Mod{T_2}$, where $T_1>T_2$ are the slow and fast periods respectively, while $t$ still accounts for the discretized time in the neural dynamics. Thus, the indices $\nu$ and $\lambda$ evolve in time, during the network's dynamics, spanning the whole set of patterns $1,2,..,K_1,K_1+1\equiv 1$ and $1,2,..,K_2,K_2+1\equiv 1$ with frequencies $1/T_1$ and $1/T_2$ respectively.

The input layer is $\bm\sigma$, clamped to the signal $\bm s(t)$, \emph{i.e.} $\bm \sigma(t)^*=\bm s(t)$, in such a way that the system's cost function reads as
\begin{align}\label{FM-TAM}
    \mathcal{H}_{N}(\bm\tau, \bm\phi; \bm s(t)|\bm\eta, \bm\chi)  = -\frac{1}{N} \sum_{i,j=1}^N \lr{\sum_{\mu=1}^{K_1} \chi_i^\mu \chi_j^{\mu} \phi_i s_j(t) + \sum_{\mu=1}^{K_2 }\eta_i^\mu \eta_j^\mu s_i(t) \tau_j}.
\end{align}
Thus, the layer $\phi$ perceives an overall field given by 
\begin{align}
    h_i^{\phi}(t) = \frac{1}{N} \sum_{\mu=1}^{K_1} \chi^\mu_i \sum_j \chi^\mu_j s_j(t),
\end{align}
while, similarly, the layer $\tau$ is subject to the field
\begin{align}
    h_i^{\tau}(t) = \frac{1}{N} \sum_{\mu=1}^{K_2} \eta^\mu_i \sum_j \eta^\mu_j s_j(t).
\end{align}
The sketch of this setting is shown in Fig.~\ref{fig:f_mod_schema}. 

\begin{figure}
    \centering
    \includegraphics[width=0.8\linewidth]{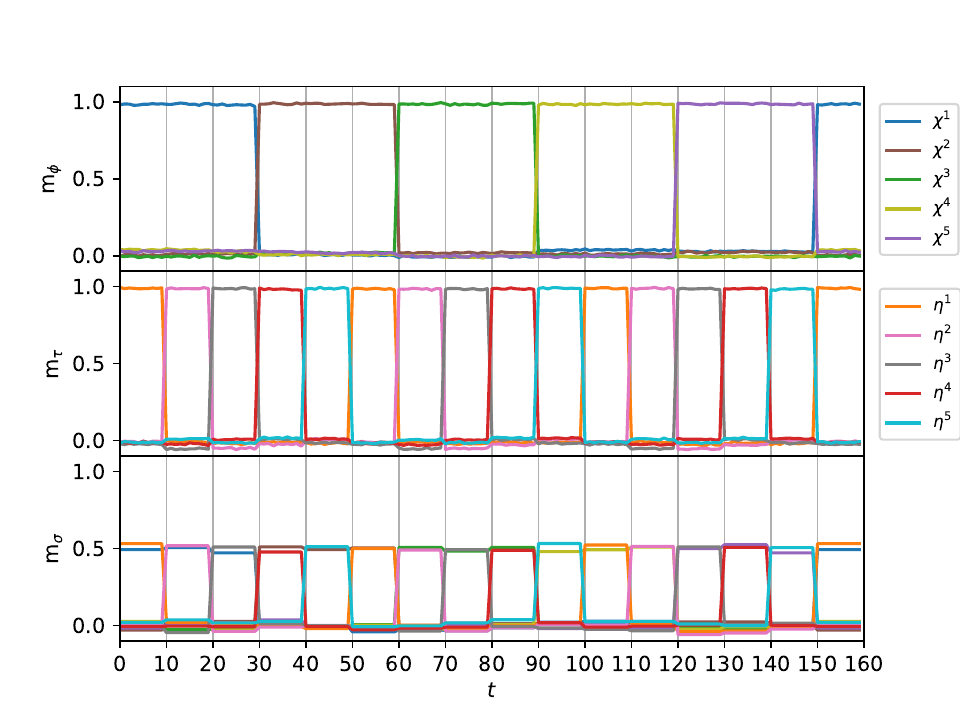}
    \caption{Frequency modulation operated by the TAM. First we mix two sequences of patterns generated at different frequency: the frequency ratio between the \emph{slow} sequence, evolving at frequency $f_1=1/T_1$ (whose signal component is retrieved in the $\tau$ layer) and the \emph{fast} sequence, evolving at frequency $f_2=1/T_2$ (whose signal component is retrieved in the $\phi$ layer) is $\frac{f_1}{f_2} = 1/3$. Then we present this composed signal to the layer $\phi$ that works as a clamped input layer: the task the TAM has to solve is to disentangle the two series and return all the temporally-ordered patterns involved in the input. This is achieved layer per layer as shown in the plots: the two upper plots report the behavior of the output layers, while the lower plot depicts the activity of the input layer (hence, it shows the signal that is provided to the TAM that has to be resolved). Obviously the upper panel returns the base-band while the middle panel returns the fast sequence, providing a successful  performance in disentanglement of patterns (or decoding) by the TAM network.\\
    In these simulations the network has three-layers all built of by $N=M=L=1000$ neurons and a total amount of patterns to handle given by $K=K_1+K_2=10$ split into $ \{\bm \chi^{\mu}\}_{\mu=1}^{K_1}$ and $ \{\bm \eta^{\mu}\}_{\mu=1}^{K_2}$ with $K_1 = K_2 =5$.}
    \label{fig:FM}
\end{figure}
At any fixed time step $t$, the fields that act on the layers $\bm \phi$ and $\bm \tau$ force them to retrieve only its instantaneous $\bm \chi^{\nu(t)}$ and $\bm \eta^{\lambda(t)}$ components respectively. We compute the resulting magnetization of the two layers in these two directions, $m^\phi_{\chi^\nu}$ and $m^\tau_{\eta^\lambda}$, at zero external noise level $\beta\to \infty$. With some algebra (see App. \ref{2patterns}) one can reach the following expressions for the equilibrium magnetizations given the signal \eqref{eq:s1}:
\begin{align}
    &m^\phi_{\chi^{\nu(t)}} = \erf\lr{\frac{1}{2\sqrt{\gamma_1}}},\\
    &m^\tau_{\eta^{\lambda(t)}} = \erf\lr{\frac{1}{2\sqrt{\gamma_2}}}.
\end{align}
In App.~\ref{2patterns} a similar analysis is conducted for the signal \eqref{eq:s2}, yielding
\begin{align}
    &m^\phi_{\chi^{\nu(t)}} = \erf\lr{\frac{1}{2\sqrt {2 \gamma_1}}},\\
    &m^\tau_{\eta^{\lambda(t)}} = \erf\lr{\frac{1}{2\sqrt{ 2\gamma_2}}}.
\end{align}

Therefore, as long as $\gamma_1$ is not too large, both magnetizations remain close to $1$ and the network can successfully reconstruct the patterns, see Fig.~\ref{fig:FM} for an example. As discussed in the App.~\ref{2patterns} amongst the two cases considered, i.e. the linear \eqref{eq:s1} and the non-linear \eqref{eq:s2} combinations,  the most challenging is the latter, due to the convention $\textrm{sign}(0)=1$ which introduces a bias in the signal.

\section{Conclusions} \label{sec:conclusions}

\setcounter{equation}{0}
\renewcommand\theequation{B.\arabic{equation}}

In this paper we generalized Kosko's bidirectional associative memories (BAM) toward a three-directional neural network (TAM), and we studied its information processing  capabilities at the replica symmetric level of description: these networks implement the so-called {\em generalized Hebbian learning rule} that allows them to handle multiple patterns at once. At difference with auto-associative neural networks (that handle one pattern per time), the possibility to simultaneously retrieve several patterns enriches the computational capabilities of these networks. 
\newline
For static memories, beyond {\em generalized pattern recognition}, they can perceive as input a mixture of patterns and return as output the original patterns that gave rise to the mixture, a task that we called  {\em pattern disentanglement}.  
\newline
Furthermore, in a dynamic environment, they can retrieve and disentangle mixtures of patterns evolving periodically in time with the same frequency or with different ones. Should the pattern sequences  alternate with different frequencies (such that we can provide as input to the network  a quickly varying sequence of patterns combined to a slowly varying sequence of patterns), recognition and disentangling still hold and the slow sequence can be used as base-band signal while the fast one as the information carrier, providing a structured variation of {\em frequency modulation}.
\newline
Clearly, this picture holds solely if each layer has to cope with its own set of patterns and as long as the control parameters are suitably set. This quest highlights once more how the role of statistical mechanics in understanding information processing systems. However, if the mixture to disentangle is a genuine Hopfield spurious state (namely we do not know a priori the pattern-layer relation) the disentangling task becomes totally unsupervised and the present network does not suffice any longer. Then, we will need networks that enjoy  both auto-associative as well as hetero-associative properties as we deepen in a forthcoming paper \cite{ABCRT2025}.
\newline
{\em En route} for conclusions:
\newline
For the Machine Learning Community reader, practical applications of these generalized hetero-associative networks can be so broad that is almost pointless to pick up one out of many. We simply remark that in video as well as in audio signal processing we often deal with mixtures of three primary inputs: in the former, the {\em RGB} (red, green, blue) composition of colours is the standard for obtaining coloured images, in the latter elementary chords are again build of by triplets of notes (as, e.g. the {\em C-major} chord is the 3-pattern mixture of the notes {\em C, E} and {\em G}).
\newline
For the Neuroscience Community reader, it is interesting to address in our model the {\em connectionist vs localist} dilemma about information processing in biological neural networks: oversimplifying, the two possible scenarios regarding how these networks store and represent patterns are the {\em parallel distributed processing} approach (where information  storage is spread over the whole network and there is no room for neurons covering different roles\footnote{We note that this perspective is the cornerstone of Connectionist Theories (which, in turn, are the main inspiration for a statistical mechanical treatment of neural networks since the original AGS theory for pairwise Hebbian learning \cite{Amit}).}), and the {\em localist perspective}, which instead assumes the existence of cells highly selective to entire  patterns (i.e., the grandmother cells \cite{BowRev}) and typically involves a structural organization of the neural network \cite{NeuralSyntax,NetOrganization,Progress2002}.  There are pros and cons to both theories: there is empirical evidence toward rejecting the claim of condensed knowledge, hence in favor of connectionist theories, yet there is also empirical evidence that support the localist hypothesis\footnote{e.g. visual neurons in the inferior temporal cortex and neurons in the upper stages of the ventral visual pathway respond to complex image of the monkey, firing selectively to hands and faces  \cite{Proof6,Proof7,Proof8,Proof9,Proof10}, further grandmother cells could easily explain synfire chains \cite{Synfire2012}, that are patterns of highly correlated firing cells \cite{Abeles1982}).}. From our, statistical mechanical driven, modelling picture we can rely upon the integral representation of the partition functions and look at the dummy integration variables as hidden neurons: remarkably, in this analogy, these hidden neurons are grandmother cells hence, while in the direct formulation (with solely visible neurons) the connectionist picture seems the more natural, from the integral formulation the localist paradigm seems more natural. Whatever the correct one (if any), they just play as two faces of the same coin in our context: the present equivalence generalizes the  already known duality of representation of Hebbian networks (i.e. Hopfield-like models \cite{Hopfield}), prototype for biological learning, in terms of restricted Boltzmann machines \cite{ContrDiv} (building blocks of modern deep architectures in Machine Learning) beyond the shallow limit \cite{BarraEquivalenceRBMeAHN,Aquaro-EPL,Agliari-Emergence,Remi2,Remi1}.

\newpage

\setcounter{equation}{0}
\renewcommand\theequation{A.\arabic{equation}}

\appendix

\section{Guerra interpolation for replica symmetric quenched free energy}\label{Appendix-Guerra}
In this Appendix we exploit Guerra's interpolation \cite{Guerra2,Fachechi1} to express the quenched free energy explicitly as a function of the control and order parameters of the theory. The plan is to use one-parameter interpolation, $t \in (0,1)$ ({\em vide infra}) and, as standard in high-storage investigations \cite{Amit,Coolen}, we assume that a finite number of patterns (actually just one per layer, with no loss of generality) is retrieved, say the triplet $(\bm\xi^1,\bm\eta^1,\bm\chi^1)$, and these patterns play as the {\em signal}, while all the remaining ones (i.e., those with labels $\nu \neq 1$) play as {\em quenched noise} against the  retrieving process.
\newline
For completeness, we recall the complete list of observables:
\begin{equation}
    \begin{array}{llllll}
         m_{\xi^\mu}^\sigma = \dfrac{1}{N}\SOMMA{i=1}{N}\xi_i^\mu\sigma_i\,,\;&& m_{\eta^\mu}^\tau = \dfrac{1}{M}\SOMMA{i=1}{M}\eta_i^\mu\tau_i\,,\;&& m_{\chi^\mu}^\phi = \dfrac{1}{L}\SOMMA{i=1}{L}\chi_i^\mu\phi_i\,,
         \\\\
         q^{\sigma}_{ab} = \dfrac{1}{N}\SOMMA{i=1}{N}\sigma_i^{(a)}\sigma_i^{(b)}\,,\;&& q^\tau_{ab} = \dfrac{1}{M}\SOMMA{i=1}{M}\tau_i^{(a)}\tau_i^{(b)}\,,\;&&q^\phi_{ab} = \dfrac{1}{L}\SOMMA{i=1}{L}\phi_i^{(a)}\phi_i^{(b)}\,,
         \\\\
         p^{z}_{ab} = \dfrac{1}{K-1}\SOMMA{\mu>1}{K}{z_\mu^\dag}^{(a)}{z_\mu^\dag}^{(b)}\,,\;&& p^{s^\dag}_{ab} = \dfrac{1}{K-1}\SOMMA{\mu>1}{K}{s_\mu^\dag}^{(a)}{s_\mu^\dag}^{(b)}\,,\;&&p^{w^\dag}_{ab} = \dfrac{1}{K-1}\SOMMA{\mu>1}{K}{w_\mu^\dag}^{(a)}{w_\mu^\dag}^{(b)}\,
         \\\\
         p^{z}_{ab} = \dfrac{1}{K-1}\SOMMA{\mu>1}{K}z_\mu^{(a)}z_\mu^{(b)}\,,\;&& p^s_{12} = \dfrac{1}{K-1}\SOMMA{\mu>1}{K}s_\mu^{(a)}s_\mu^{(b)}\,,\;&&p^w_{12} = \dfrac{1}{K-1}\SOMMA{\mu>1}{K}w_\mu^{(a)}w_\mu^{(b)}\,.
    \end{array}
\end{equation}

As we deal with averages of observables, it is useful to introduce them as follows.
\newline
Given a function $f(\sigma,\tau,\phi)$, depending on the neuronal configuration $(\sigma,\tau,\phi)$, 
the Boltzmann average, namely the average over the distribution \eqref{BGmeasure}, is denoted as $\omega( f(\sigma,\tau,\phi))$ and defined as
$$
 \ \omega(f(\sigma,\tau,\phi))= \frac{\sum_{\{\sigma \}}^{2^N} \sum_{\{\tau \}}^{2^M} \sum_{\{\phi \}}^{2^L} f(\sigma,\tau,\phi) e^{-\beta H_N(\boldsymbol{s}|\boldsymbol{\eta})}}{\sum_{\{\sigma \}}^{2^N} \sum_{\{\tau \}}^{2^M} \sum_{\{\phi \}}^{2^L} e^{-\beta H_N(\boldsymbol{s}|\boldsymbol{\eta})}}.
$$
Further, given a function of the weights\footnote{It is not a mistake to call the patterns as {\em weights} because, in the integral representation of the partition function, they play exactly that role: see Eq. \eqref{Integralista}.} $g(\xi, \eta, \chi)$ depending on the realization of the $K$ triplets of patterns, we introduce the quenched average, namely the average over the Rademacher distributions related to their generation, that is denoted as $\mathbb E [ g(\xi, \eta, \chi)]$ or as $\langle g(\xi, \eta, \chi) \rangle$ according to the context, and it is defined as
\begin{align}
\label{eq:mapsexpectation}
    \mathbb{E} [ g(\xi, \eta, \chi) ] \equiv \langle g(\xi, \eta, \chi)\rangle = \mathbb{E}_{\xi}\mathbb{E}_{\eta}\mathbb{E}_{\chi}  \mathcal{P}(\xi, \eta, \chi) g(\xi, \eta, \chi).
\end{align}
This definition of the quenched average follows from the assumption that the patterns are statistically independent and therefore the expectation factorizes over the neurons $i=1,..,N$ and over the patterns $\mu=1,..,K$.
\newline
Finally, we denote with the brackets $\langle \cdot \rangle$ the average over both the Boltzmann-Gibbs distribution and the realization of the patterns, that is
$$
\langle f((\sigma,\eta,\phi)|(\xi, \eta, \chi)) \rangle = \mathbb E [\omega(f((\sigma,\eta,\phi)|(\xi, \eta, \chi)))].
$$ 
Under the replica symmetric ansatz, we assume that the probability distributions  of the order parameters become Dirac deltas in the thermodynamic limit, i.e.
\begin{equation}
    \begin{array}{lllll}
        \P(m_{\xi^1}^\sigma) \xrightarrow[]{N\to\infty} \delta(m_{\xi^1}^\sigma-\m)\,,\;&& \P(m_{\eta^1}^\tau)  \xrightarrow[]{N\to\infty} \delta(m_{\eta^1}^\tau-\n)\,,\;&&\P(m_{\chi^1}^\phi)  \xrightarrow[]{N\to\infty} \delta(m_{\chi^1}^\phi-\llm)\,,
         \\\\
         \P(q^{\sigma}_{ab})  \xrightarrow[]{N\to\infty} \delta(q^\sigma_{ab}-\qsigma)\,,\;&& \P(q^\tau_{ab})  \xrightarrow[]{N\to\infty} \delta(q^\tau_{ab}-\qtau)\,,\;&&\P(q^\phi_{ab})  \xrightarrow[]{N\to\infty} \delta(q^\phi_{ab}-\qphi)\,,
         \\\\
         \P(p^{z}_{ab}) \xrightarrow[]{N\to\infty} \delta(p^{z}_{ab}-\bar p^z)\,,\;&& \P(p^{s^\dag}_{ab})  \xrightarrow[]{N\to\infty} \delta(p^{s}_{ab}-\bar p^s)\,,\;&&\P(p^{w^\dag}_{ab})  \xrightarrow[]{N\to\infty} \delta(p^{w}_{ab}-\bar p^w)\,
         \\\\
         \P(p^{z^\dag}_{ab})  \xrightarrow[]{N\to\infty} \delta(p^{z^\dag}_{ab}-\pdagz)\,,\;&& \P(p^{s^\dag}_{ab})  \xrightarrow[]{N\to\infty} \delta(p^{s^\dag}_{ab}-\pdags)\,,\;&&\P(p^{w^\dag}_{ab})  \xrightarrow[]{N\to\infty} \delta(p^{w^\dag}_{ab}-\pdagw)\,. 
    \end{array}
\end{equation}
hence the expectations of the order parameters collapse on these values in this asymptotic limit where the theory is worked out, that is, calling $x$ a generic order parameter, $\lim_{N \to \infty}\langle x(\sigma,\tau,\phi)\rangle = \bar{x}$.
\newline
\newline
The idea to solve for the free energy in this framework is to introduce an interpolating parameter $t\in[0,1]$ and an interpolating free energy $\mathcal A(t)$ such that, when $t=1$, the interpolating free energy recovers the free energy of the original model, i.e., $\mathcal A(t=1) = A^{\a,\b,\c}_{\alpha, \theta, \gamma}(\beta)$, while, when $t=0$, the interpolating free energy recovers the free energy of an ``easy'' one-body system where neurons interact with a suitably-constructed external field but their activity is no longer affected by the state of the other neurons: the external field must be constructed to best mimic -at least at the lowest order statistics- the internal field originally produced by the true neurons. Furthermore, we can introduce auxiliary functions and constant in the one-body expression (i.e. $A(t=0)$) so to have the freedom to fix them {\em a fortiori} in order to simplify calculations and impose replica symmetry.  
\\
As a result, the main theorem we use in this section is just the fundamental theorem of calculus  that plays as the natural bridge between these two extrema and gives rise to the following sum-rule:  
\begin{eqnarray}
\label{InterMilan}
    A^{\a,\b,\c}_{\alpha, \theta, \gamma}(\beta)&=&\mathcal A(t=1) = \mathcal A(t=0) + \int_0^1 ds \left. \left[\frac{b}{dt} \mathcal A(t)\right] \right \vert_{t=s}
    \\ \label{Mortaccitua}
    \mathcal A(t) &=& \lim_{N\to\infty} \frac{1}{N} \mathbb E_\eta  \ln \mathcal Z(t),
 \end{eqnarray}
where $\mathcal Z(t)$ is the Guerra's interpolating partition function defined hereafter.
\\
Once introduced real valued auxiliary functions $\psi$ (that play as {\em ad hoc} fields to reproduce the neural interactions effectively) and the set of constants $C$ (that properly tune the variances of the distributions of the real-valued hidden grandmother neurons), the Guerra's interpolating partition function reads as
\begin{equation}\label{Integralista}
    \begin{array}{lll}
         \mathcal Z(t)=\SOMMA{\{\sigma\},\{\tau\},\{\phi\}}{2^N,2^M,2^{L}} \exp\left[t\beta\left(\a\sqrt{NM}m^\sigma_{\xi^1}m^\tau_{\xi^1}+\b\sqrt{NL}m^\sigma_{\xi^1}m^\phi_{\xi^1}+\c\sqrt{L M}m^\phi_{\xi^1}m^\tau_{\xi^1}\right)\right.
    \\\\
         \left.+(1-t)\left(\sqrt{NM}(\psi_m^{(1)}m^\sigma_{\xi^1}+\psi_n^{(1)}m^\tau_{\xi^1})+\sqrt{NL}(\psi_m^{(2)}m^\sigma_{\xi^1}+\psi_l^{(1)}m^\phi_{\xi^1})+\sqrt{L M}(\psi_l^{(2)}m^\phi_{\xi^1}+\psi_n^{(2)}m^\tau_{\xi^1})\right)\right]\times
    \\\\
         \times\displaystyle\int\mathcal{D}(z\z s\s w\w)\exp\left[\sqrt{t}\sqrt{\dfrac{\beta}{2N}}\SOMMA{\mu>1}{K}\SOMMA{i=1}{N}\xi_i^\mu\sigma_i z_\mu+\sqrt{t}\sqrt{\dfrac{\beta}{2M}}\SOMMA{\mu>1}{K}\SOMMA{j=1}{M}\eta_j^\mu\tau_j \s_\mu+\sqrt{t}\sqrt{\dfrac{\beta}{2\N}}\SOMMA{\mu>1}{K}\SOMMA{k=1}{L}\chi_k^\mu\phi_k \w_\mu\right.
    \\\\
         +\dfrac{(1-t)}{2} \left(\bar C_z\SOMMA{\mu>1}{K} \left(\z_\mu\right)^2+C_z\SOMMA{\mu>1}{K} \left(z_\mu\right)^2+ \bar C_s\SOMMA{\mu>1}{K} \left(\s_\mu\right)^2+ C_s\SOMMA{\mu>1}{K} \left(s_\mu\right)^2+ \bar C_w\SOMMA{\mu>1}{K} \left(\w_\mu\right)^2+ C_w\SOMMA{\mu>1}{K} \left(w_\mu\right)^2\right)
    \\\\
         + \sqrt{1-t}\left( B_z\SOMMA{\mu=1}{K}J_\mu^z z_\mu+\bar B_z\SOMMA{\mu=1}{K}\bar J_\mu^z \z_\mu+ B_s\SOMMA{\mu=1}{K}J_\mu^s s_\mu+\bar B_s\SOMMA{\mu=1}{K}\bar J_\mu^s \s_\mu+ B_w\SOMMA{\mu=1}{K}J_\mu^w w_\mu+\bar B_w\SOMMA{\mu=1}{K}\bar J_\mu^w \w_\mu\right)
  \\\\
        + \sqrt{1-t}\left( A_\sigma\SOMMA{i=1}{N}Y_i^\sigma \sigma_i+ A_\tau\SOMMA{k=1}{M}Y_k^\tau \tau_k+ A_\phi\SOMMA{j=1}{L}Y_j^\phi \phi_j\right)\Bigg]
    \end{array}
\end{equation}
Note that $\mathcal Z(t=1)$ does coincide with the partition function of the original TAM network, while $\mathcal Z(t=0)$ is certainly analytically valuable as it is a linear combination of one-body models, whose underlying probabilistic measure is trivially factorized by definition.
\newline
In the following, if not otherwise specified, we refer to the generalized averages simply as $\langle \cdot \rangle$ in order to lighten the notation.\\
By a glance at \eqref{InterMilan} we see that we have to evaluate the Cauchy condition  $\mathcal A(t=0)$ (that is straightforward as in $t=0$ the Gibbs measure is factorized over the neural activities) and integrate the t-streaming, that is the derivative  $\frac{d\mathcal A(t)}{dt}$, over the interval $t \in (0,1)$. 
\\
The explicitly calculation of the $t$-derivative of the interpolating free energy introduced in eq. \eqref{InterMilan} works by brute force thus we report directly our findings.
\begin{equation}
    \begin{array}{lllll}
    \partial_t \mathcal{A}(t)=\dfrac{1}{3}\left(\dfrac{1}{\sqrt{NL}}+\dfrac{1}{\sqrt{ML}}+\dfrac{1}{\sqrt{NM}}\right)\Bigg[\beta\left(\a\sqrt{NM}m^\sigma_{\xi^1}m^\tau_{\xi^1}+\b\sqrt{NL}m^\sigma_{\xi^1}m^\phi_{\xi^1}+\c\sqrt{LM}m^\phi_{\xi^1}m^\tau_{\xi^1}\right)
    \\\\
         \left.-\left(\sqrt{NM}(\psi_m^{(1)}m^\sigma_{\xi^1}+\psi_n^{(1)}m^\tau_{\xi^1})+\sqrt{NL}(\psi_m^{(2)}m^\sigma_{\xi^1}+\psi_l^{(1)}m^\phi_{\xi^1})+\sqrt{LM}(\psi_l^{(2)}m^\phi_{\xi^1}+\psi_n^{(2)}m^\tau_{\xi^1})\right)\right]
    \\\\
         \dfrac{\beta}{4}K(p^z_{11}-\q^\sigma p^z_{12})+\dfrac{\beta}{4}K(\ps_{11}-\q^\tau \ps_{12})+\dfrac{\beta}{4}K(\pw_{11}-\q^\phi \pw_{12})
    \\\\
         -\dfrac{K}{2} \left(\bar C_z \pz_{11}+C_z p^z_{11}+ \bar C_s \ps_{11}+ C_s p^s_{11}+ \bar C_w \pw_{11}+ C_w p_{11}^w\right)
    \\\\
         -\dfrac{K}{2}\left( B_z^2 (p_{11}^z-p_{12}^z)+\bar B_z^2 (\pz_{11}-\pz_{12})+ B_s^2 (p_{11}^s-p_{12}^s)+\bar B_s^2 (\ps_{11}-\ps_{12})+ B_w^2 (p_{11}^w-p_{12}^w)+\bar B_w^2 (\pw_{11}-\pw_{12})\right)
  \\\\
        -\dfrac{1}{2}\left( N A_\sigma^2(1-q^\sigma)+ M A_\tau^2(1-q^\tau)+L A_\phi^2(1-q^\phi)\right)\Bigg]
    \end{array}
\end{equation}
We see that, if we fix the values of the constants and the auxiliary field as follow
\begin{equation}
    \begin{array}{lllll}
         \bar C_z =-\bar B_z=0 &&C_s =-B_s=0 && C_w = -B_w=0
         \\\\
         C_z = \dfrac{\beta}{2}(1-\qsigma) && \bar C_s = \dfrac{\beta}{2}(1-\qtau) && \bar C_w = \dfrac{\beta}{2}(1-\qphi)
         \\\\
         B_z^2=\dfrac{\beta}{2}\qsigma && \bar B_s^2=\dfrac{\beta}{2}\qtau && \bar B_w^2=\dfrac{\beta}{2}\qphi
         \\\\
         A_\sigma^2=\dfrac{\beta}{2}\dfrac{K}{N}\bar p^z && A_\tau^2=\dfrac{\beta}{2}\dfrac{K}{M}\pdags && A_\phi^2=\dfrac{\beta}{2}\dfrac{K}{\N}\pdagw
         \\\\
         \psi_m^{(1)}=\a\beta \n && \psi_n^{(1)}=\a\beta \m && \psi_l^{(1)}=\b\beta \m
         \\\\
         \psi_m^{(2)}=\b\beta \llm && \psi_n^{(2)}=\c\beta \llm && \psi_l^{(2)}=\c\beta \n
    \end{array}
\end{equation}
we get the final expression in the thermodynamic limit of the derivative of the interpolating free energy that reads as
\begin{equation}\small
\label{eq:straming_computed}
    \begin{array}{lllll}
    \partial_t \mathcal{A}(t)&=&-\beta\varsigma\left[\a\theta^{-1}\m\n+\b\alpha^{-1}\m\llm+\c(\alpha\theta)^{-1}\llm\n\right]
    \\\\
    &&-\dfrac{\varsigma}{4}\left(\alpha+\theta+\alpha\theta\right)\left[ \beta\bar p^z(1-\qsigma)+ \beta\pdags(1-\qtau)+\beta\pdagw(1-\qphi)\right]
    \end{array}
\end{equation}
where we use $\varsigma = \dfrac{1}{3}(\alpha+\theta+\alpha\theta)$\,.

Now, in order to fulfil the sum rule stemmed by the fundamental theorem of calculus for the interpolating free energy -see eq. \eqref{InterMilan}- we have to evaluate the Cauchy condition $\mathcal A(t=0)$: by construction, as we are interpolating the original model with a linear combination of one-body contributions, such condition can be explicitly evaluated, again by brute force, and gives rise to
$$
\mathcal A(t=0) = \lim_{N \to \infty} \frac{1}{N}\ln \mathbb{E}\mathcal{Z}(t=0),
$$
where the interpolating partition function, evaluated at $t=0$, reads as
\begin{equation}
    \begin{array}{lllll}
         \mathcal{Z}(t=0)=\prod\limits_{i=1}^{N}2\cosh\left[\xi_i^\mu\psi_m^{(1)}\sqrt{\dfrac{M}{N}}+\xi_i^\mu\psi_m^{(2)}\sqrt{\dfrac{\N}{N}}+A_\sigma Y_i^\sigma \right]\prod\limits_{k=1}^{M} 2\cosh\left[\eta_k^\mu\psi_n^{(1)}\sqrt{\dfrac{N}{M}}+\eta_k^\mu\psi_n^{(2)}\sqrt{\dfrac{\N}{M}}+A_\tau Y_k^\tau \right]\times
         \\\\
         \times\prod\limits_{j=1}^{\N}2\cosh\left[\chi_j^\mu\psi_l^{(1)}\sqrt{\dfrac{N}{\N}}+\chi_j^\mu\psi_l^{(2)}\sqrt{\dfrac{M}{\N}} +A_\phi Y_j^\phi \right]\times\prod\limits_{\mu>1}^{K}\displaystyle\int \dfrac{b^6\bm X_\mu }{(\sqrt{2\pi})^6}\exp\Bigg[-\dfrac{1}{2} \bm X_\mu^T \bm C \bm X_\mu + \bm X_\mu^T \bm B_\mu\Bigg]
    \end{array}
\end{equation}
where we have set
\begin{equation}
\begin{array}{llll}
     \bm X_\mu= \begin{pmatrix}
        Re(z_\mu)
        \\
       Im(z_\mu)
        \\ 
        Re(s_\mu)
        \\
        Im(s_\mu)
        \\ 
        Re(w_\mu)
        \\
        Im(w_\mu)
    \end{pmatrix}\,,\;&& \bm B_\mu= \begin{pmatrix}
        B_z J_\mu^z + \bar B_z \bar J_\mu^z
        \\
        i(B_z J_\mu^z - \bar B_z \bar J_\mu^z)
        \\ 
         B_s J_\mu^s + \bar B_s \bar J_\mu^s
        \\
        i(B_s J_\mu^s - \bar B_s \bar J_\mu^s)
        \\ 
         B_w J_\mu^w + \bar B_w \bar J_\mu^w
        \\
        i(B_w J_\mu^w - \bar B_w \bar J_\mu^w)
    \end{pmatrix}
\end{array}
\end{equation}

\begin{equation} \label{eq:A14}
    \bm C= \begin{pmatrix}
        1-\bar C_z -C_z & i(\bar C_z -C_z) & -\a/2&-i\a/2&-\b/2&-i\b/2
        \\
        i(\bar C_z -C_z) &1+\bar C_z +C_z & i\a/2&-\a/2&ib/2&-\b/2
        \\ 
        -\a/2&ia/2&1-\bar C_s -C_s & i(\bar C_z -C_z)&-\c/2&-i\c/2
        \\
        -i\a/2&-\a/2&i(\bar C_s -C_s) &1+\bar C_z +C_z&-i\c/2&\c/2
        \\ 
        -\b/2&ib/2&-\c/2&-i\c/2&1-\bar C_w -C_w & i(\bar C_w -C_w)
        \\
        -i\b/2&-\b/2&-i\c/2&\c/2&i(\bar C_w -C_w) &1+\bar C_w +C_w
    \end{pmatrix}
\end{equation}
Finally, computing the Gaussian integrals, we get
\begin{equation}
    \begin{array}{lllll}
         \mathcal{Z}(t=0)=\prod\limits_{i=1}^{N}2\cosh\left[\xi_i^\mu\psi_m^{(1)}\sqrt{\dfrac{M}{N}}+\xi_i^\mu\psi_m^{(2)}\sqrt{\dfrac{\N}{N}}+A_\sigma Y_i^\sigma \right]\prod\limits_{k=1}^{M} 2\cosh\left[\eta_k^\mu\psi_n^{(1)}\sqrt{\dfrac{N}{M}}+\eta_k^\mu\psi_n^{(2)}\sqrt{\dfrac{L}{M}}+A_\tau Y_k^\tau \right]\times
         \\\\
         \times\prod\limits_{j=1}^{\N}2\cosh\left[\chi_j^\mu\psi_l^{(1)}\sqrt{\dfrac{N}{\N}}+\chi_j^\mu\psi_l^{(2)}\sqrt{\dfrac{M}{\N}} +A_\phi Y_j^\phi \right]\times\left(\mathrm{det}\bm C\right)^{-K/2}\prod\limits_{\mu>1}^{K}\exp\Bigg[\dfrac{1}{2} \bm B_\mu^T \bm C^{-1} \bm B_\mu \Bigg],
    \end{array}
\end{equation}
thus the one-body contribution to the interpolating free energy is
\begin{equation}
\label{eq:one_body_computed}
    \begin{array}{lllll}
         \mathcal{A}(t=0)&=&\varsigma\,\mathbb E\left\{\log 2\cosh\left[\xi^\mu\left(\psi_m^{(1)}\dfrac{1}{\theta}+\psi_m^{(2)}\dfrac{1}{\alpha}\right)+A_\sigma Y^\sigma \right]\right\}
         \\\\
         &&+\dfrac{\varsigma}{\theta^2}\mathbb E\left\{\log 2\cosh\left[\eta^\mu\theta\left(\psi_n^{(1)}+\psi_n^{(2)}\dfrac{1}{\alpha}\right)+A_\tau Y^\tau \right]\right\}
         \\\\
         &&+\dfrac{\varsigma}{\alpha^2}\mathbb{E}\left\{\log 2\cosh\left[\chi^\mu\alpha\left(\psi_l^{(1)}+\psi_l^{(2)}\dfrac{1}{\theta}\right) +A_\phi Y^\phi \right]\right\}
         \\\\
         &&-\dfrac{\gamma\varsigma}{2}\log\left[\mathrm{det}\bm C\right]+\dfrac{\gamma\varsigma}{2}\mathbb E\Bigg[ \bm B_\mu^T \bm C^{-1} \bm B_\mu \Bigg].
    \end{array}
\end{equation}

Merging these two expressions, that is the integral of the $t$-streaming (Eq.~\eqref{eq:straming_computed}) and the Cauchy condition (Eq.~\eqref{eq:one_body_computed}), in Eq.\eqref{InterMilan} under the replica symmetric assumption,  we get the explicit expression of the quenched free energy in terms of control and order parameters of the theory as planned: this reads as 

\begin{equation}\label{eq:fRS}
    \begin{array}{lllll}
         A_{\alpha,\theta,\gamma}^{\a,\b,c}(\beta)&=&\varsigma\,\mathbb E\left\{\log 2\cosh\left[\beta\xi^1\left( \n\dfrac{\a}{\theta}+ \llm\dfrac{\b}{\alpha}\right)+ Y^\sigma \sqrt{\dfrac{\beta}{2}\gamma\bar p^z}\right]\right\}
         \\\\
         &&+\theta^{-2}\varsigma\,\mathbb E\left\{\log 2\cosh\left[\beta\eta^1\theta\left( \a\m+ \llm\dfrac{\c}{\alpha}\right)+ Y^\tau \sqrt{\dfrac{\beta}{2}\theta^2\gamma\pdags}\right]\right\}
         \\\\
         &&+\alpha^{-2}\varsigma\,\mathbb{E}\left\{\log 2\cosh\left[\beta\chi^1\alpha\left( \b\m+ \n\dfrac{\c}{\theta}\right) + Y^\phi\sqrt{\dfrac{\beta}{2}\alpha^2\gamma\pdagw} \right]\right\} -\dfrac{\gamma\varsigma}{2}\log\left[\mathrm{det}\bm C\right]
    \\\\
    && +\gamma\varsigma\beta^3\a\b\c\dfrac{  \qsigma(1 - \qtau)(1 - \qphi)     +(1 - \qsigma) (1 - \qtau)\qphi    + (1 - \qsigma) \qtau   (1 - \qphi) }{\mathrm{det} \bm C}
    \\\\
    &&
    +\dfrac{\gamma\varsigma}{2}\dfrac{\beta^2 \a^2  (1 - \qsigma) \qtau + \beta^2 \a^2 \qsigma(1 - \qtau)  + 
 \beta^2 \c^2 \qtau(1 - \qphi )  }{\mathrm{det} \bm C}
 \\\\
 && +\dfrac{\gamma\varsigma}{2}\dfrac{\beta^2 \c^2 (1 - \qtau) \qphi + 
 \beta^2 \b^2 (1 - \qphi) \qsigma + \beta^2 \b^2(1 - \qsigma) \qphi}{\mathrm{det} \bm C}
    \\\\
        && -\beta\varsigma\left[\a\theta^{-1}\m\n+\b\alpha^{-1}\m\llm+\c(\alpha\theta)^{-1}\llm\n\right]
    \\\\
        &&-\dfrac{\gamma\varsigma}{4}\left[ \beta\bar p^z(1-\qsigma)+ \beta\pdags(1-\qtau)+\beta\pdagw(1-\qphi)\right]
    \end{array}
\end{equation}
where
\begin{equation}
\small
    \mathrm{det}\bm C=1 - \Big[\beta^2\a^2(1 - \qtau) (1 - \qsigma) + 
   \beta^2\b^2(1 - \qphi)(1 - \qsigma) + 
    \beta^2\c^2 (1 - \qphi)(1 - \qtau)\Big]   - 2\a \b \c\beta^3(1 - \qphi) (1 - \qsigma)(1 - \qtau)  \, .
\end{equation}


\section{One-step magnetization} \label{app:B}
\renewcommand{\theequation}{B.\arabic{equation}}
In this section, we present the details of the computation for the evolution of Mattis magnetization after one step of neural dynamics, for the four tasks described in the main test.
\subsection{2-Patterns mixtures}\label{2patterns}
Note that the signal \eqref{eq:s1}, built of by just two patterns per time, forces the neurons $\sigma_i$ to assume values in $\{-1,0,+1\}$, while the output neurons stay binary. With this simple linear combination, all the entries where two summed patterns have opposite  values give to the neurons in the input layer zero contributions and, correspondingly, these values of $\bm \sigma$ are zeros. With this in mind we are able to explicitly compute the fields $\bm h^{\phi}$ and $\bm h^{\tau}$: at leading order in $N$ they simply read
\begin{align}
    &h_i^\phi = \frac{1}{2} \chi^\nu_i + \mathcal O (N^{-1/2}),\\
    &h_i^\tau = \frac{1}{2} \eta^\lambda_i + \mathcal O (N^{-1/2}).
\end{align}
This configuration of the two fields $\bm h^\phi$ and $\bm h^\tau$ force the two output layers $\{\bm \phi,\bm \tau\}$ to retrieve the components $\bm \chi^\nu$ and $\bm \eta^\lambda$ of the input configuration $\bm \sigma = \frac{1}{2}\lr{\bm \chi^\nu + \bm \eta^\lambda}$.  

A rather similar picture keeps  holding even in the more stringent case where the input layer $\bm \sigma$ can only take ${\pm 1}$ values, just as the other two input-layers, that is under the assumption that we provide as the input signal $\bm s(t)$ the sign function of the mixture of sequences  $\bm \sigma =\sign \lr{\bm \chi^\nu + \bm \eta^\lambda}$. The only difference with the previous case is that now, in the entries where the input combination $\chi^\nu_k + \eta^\lambda_k = 0$, the input-neurons are activated because of the convention $\sign(0)=1$.\\ \\
 
Let us deepen the differences these two inputs induce in the network by the one-step signal-to-noise analysis where we test the retrieval capabilities of the model in terms of the magnetizations $m_\phi$ and $m_\tau$ regarding the components $\bm \chi^\nu$ and $\bm \eta^\lambda$ of the signal. 
\newline
Remembering that in the $\beta \to \infty$ limit the hyperbolic tangents get error functions, preliminary we write
\begin{align}\label{Taccitua1}
    &m_\phi = \erf\lr{\frac{\mu_\phi}{\sqrt{2s_\phi}}},\\ \label{Taccitua2}
    &m_\tau = \erf\lr{\frac{\mu_\tau}{\sqrt{2 s_\tau}}}
\end{align}
with
\begin{align}
    &\mu_\phi = \mathbb E \lr{h^\phi_i \chi^\nu_i},\\
    &\mu_\tau = \mathbb E \lr{h^\tau_i \eta^\lambda_i},\\
    &s_\phi = \mathbb E \lr{\lr{h^\phi_i \chi^\nu_i}^2} - \mu_\phi^2,\\
    &s_\tau = \mathbb E \lr{\lr{h^\tau_i \eta^\lambda_i}^2} - \mu_\tau^2.
\end{align}
\begin{itemize}
\item In the simplest scenario of linear combination of patterns, $\bm \sigma =\frac{1}{2} \lr{\bm \chi^\nu + \bm \eta^\lambda}$ and we have
\begin{align}
    &\mu_\phi = \frac{1}{2},\\
    &\mu_\tau  = \frac{1}{2},\\
    &s_\phi = \frac{1}{2}\gamma_1,\\
    &s_\tau = \frac{1}{2}\gamma_2,
\end{align}
where $\gamma_1 = K_1/N$ and $\gamma_2 = K_2/N$ are the loads pertaining to the $\bm \phi$ and $\bm \tau$ layers. Notice that, given the symmetry between the two layers, the results only depend on the loads $\gamma_1,\gamma_2$; hence, hereafter, we focus on the $\bm \phi$ layer only, while the equivalent results for $\bm \tau$ can be obtained by simply substituting $\gamma_1 \to \gamma_2$.\\
Combining the results we have
\begin{align}
    &m_\phi = \erf\lr{\frac{1}{2\sqrt{\gamma_1}}}
\end{align}
and we aim to compare these values of the output magnetizations to those stemming from the sign-function as input addressed hereafter.
\item In the non-linear case $\bm \sigma =\sign \lr{\bm \chi^\nu + \bm \eta^\lambda}$ we still have error functions rather than hyperbolic tangent (see eq.s \eqref{Taccitua1}, \eqref{Taccitua2}) but this time their arguments are different as now we have for the mean value
\begin{align}
    \mu_\phi = \frac{1}{N} \mathbb E \lr{\sum_\mu \chi^\mu_i \chi^\mu_j \chi^\nu_i \sum_{j|+}\chi^\mu_j \lr{\chi^\nu_j + \eta^\lambda_j} + \sum_\mu \chi^\mu_i \chi^\nu_i \sum_{j|-} \chi^\mu_j}
\end{align}
The summations $\sum_{j|+}$ and $\sum_{j|-}$ denote the sum restricted over the sites $j$ where the patterns $\bm \chi^\nu$ and $\bm \eta^\lambda$ have the same sign or the opposite respectively; in the first case, the values of the input $\sigma_j$ can be either $\pm 1$, according to the value of the sum $\chi_j^\nu + \eta_j^\lambda$, while in the negative case the input is always $\sigma_j=1$. Considering that in the thermodynamic limit $N\to \infty$ we have $\sum_{j|+}\chi^\mu_j \chi^\nu_j \sim \frac{N}{2} \delta^{\mu\nu} + \frac{\sqrt N}{2}(1-\delta_{\mu\nu})$ and $\sum_{j|-} \chi^\mu_j \sim 0$, we end up with
\begin{align}\label{eq:msigma}
    \mu_\phi = \frac{1}{2}.
\end{align}
Now, consider the variance term:
\begin{align}
    \mathbb E \Big[\lr{h^\phi_i \chi^\nu_i}^2 \Big] = \mathbb E \left[\lr{\frac{1}{N}\sum_\mu \chi^\mu_i \chi^\mu_j \chi^\nu_i \sum_{j|+}\chi^\mu_j \lr{\chi^\nu_j + \eta^\lambda_j} + \sum_\mu \chi^\mu_i \chi^\nu_i \sum_{j|-} \chi^\mu_j}^2\right];
\end{align}
After collecting all relevant terms in the thermodynamic limit $N\to \infty$ we end up with
\begin{align}\label{eq:ssigma}
    s_\phi = \gamma_1.
\end{align}
Notice that in this case (with a signal given by eq. \ref{eq:s1}), the variance $s_\phi$ is doubled with respect to the previous case (with a signal given by eq. \ref{eq:s2}).
Hence, combining \ref{eq:msigma} and \ref{eq:ssigma}, the magnetization reads
\begin{align}
    &m_\phi = \erf\lr{\frac{1}{2\sqrt{2\gamma_1}}}.
\end{align}
\end{itemize}
Hence the output values that the TAM returns, in the last case (with the input layer gets activated through the $\sign$ function) are globally lower than in the first case where we directly pass a linear combination in input. The non-linearity of the sign function here impoverishes the quality of the retrieval: this can be understood by noticing that the convention $\sign(0)=1$ introduces a systemic bias as an extra-source of noise in the network.
\begin{figure}
    \centering
    \includegraphics[width=0.45\linewidth]{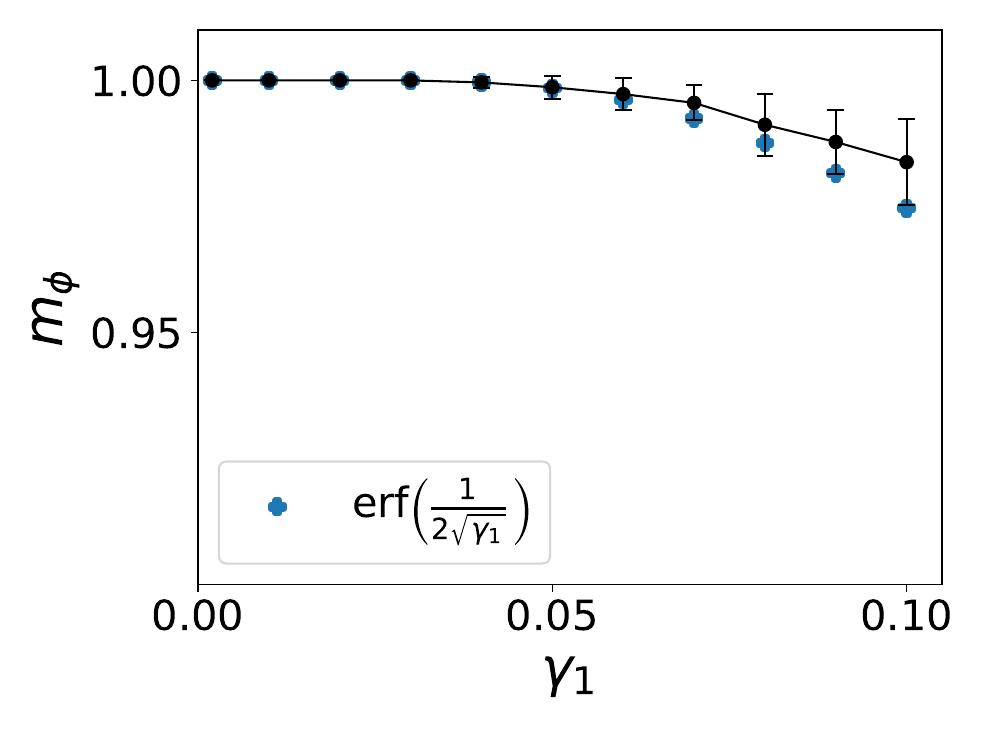}
    \includegraphics[width=0.45\linewidth]{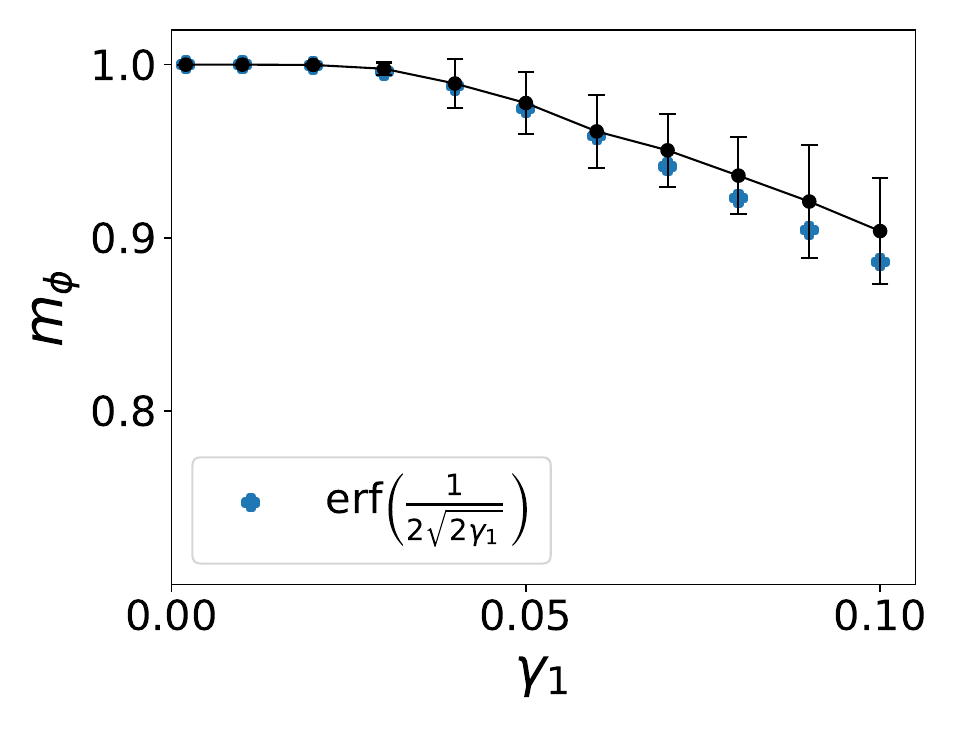}
    \caption{Left: $\phi$ magnetization after one-step update with the signal $\bm s = \frac{1}{2}(\bm \xi^\nu + \bm \eta^\lambda)$.
    Right: $\phi$ magnetization after one-step update with the signal $\bm s = \sign(\bm \xi^\nu + \bm \eta^\lambda)$.\\
    Error bars are computed as $\pm 1$ s.t.d. from experimental data (with $50$ quenches).}
    \label{fig:enter-label}
\end{figure}

\mycomment{
\begin{figure}[t]
    \centering
    \includegraphics[width=12cm]{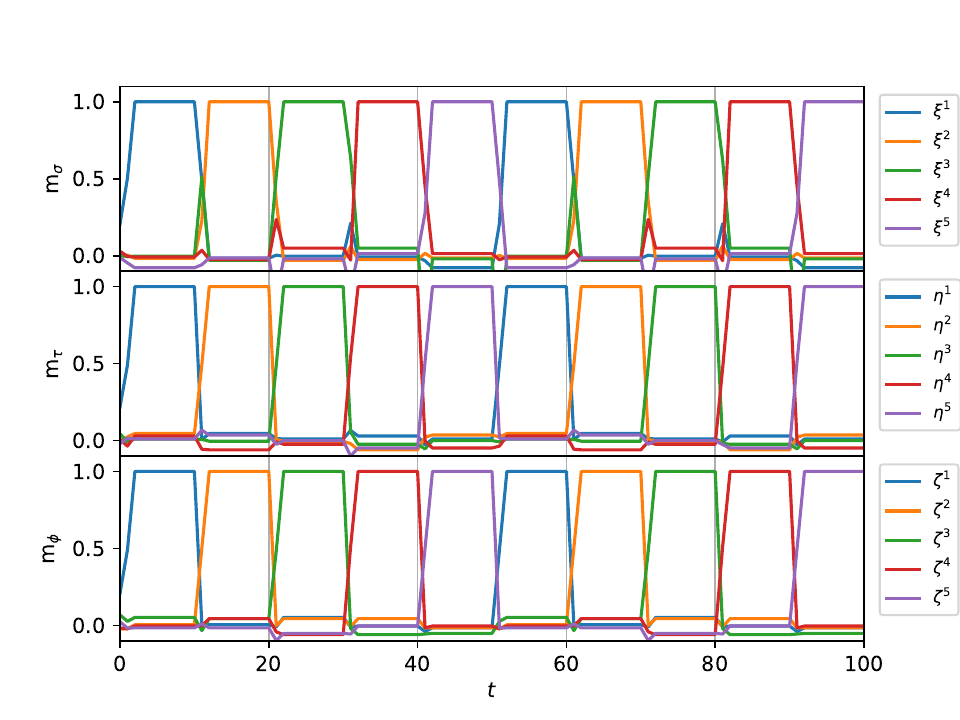}
    \caption{Frequency modulation via Detailed Balance.
    The signal is passed to the network's layer through the external field (see eq. \eqref{TAM-BilancioDettagliato-FM}) without clumping any layer.
    \newline
    In these simulations the network has three-layers all built of by $N=M=L=1000$ neurons and a total amount of patterns to handle given by $K=K_1+K_2+K_3=15$ split in $K_1 = 5$ $\xi$-patterns, $K_2 = 5$ $\eta$-patterns and $K_3 = 5$ $\chi$-patterns.}
    \label{fig:sequence-TAM}
\end{figure}
}


\subsection{Unbiased datasets}\label{Hinton}
In this section we will focus on the {\em pattern disentanglement} task.
\newline
Far from being rigorous or exhaustive, the following approach allows us to preliminary check that, by initializing the network into a mixture of three different patterns (e.g.,  $\mathrm{sign}[\bm\xi^{\mu}+\bm\eta^{\mu}+\bm\chi^{\mu}]$), the noiseless neural dynamics encoded by the master equation \eqref{dinamical} in the $\beta \to \infty$ limit, relaxes toward a configuration where each layer gets aligned with one of the constituting patterns (i.e., each of the patterns $\bm\xi^{\mu}$, $\bm\eta^{\mu}$ $\bm\chi^{\mu}$ is retrieved).
\newline  
To this purpose we define the fields acting on each neuron as 
\begin{equation}
\begin{array}{lll}
     h^\sigma_i(\bm\tau, \bm\phi)=\dfrac{1}{\sqrt{N}}\SOMMA{\mu=1}{K}\left(\dfrac{\a}{\sqrt{M}}\SOMMA{j=1}{M}\eta_j^\mu\tau_j+\dfrac{\b}{\sqrt{L}}\SOMMA{k=1}{L}\chi_k^\mu\phi_k\right)\xi_i^\mu\,,
     \\\\
     h^\tau_j(\bm\sigma, \bm\phi) =\dfrac{1}{\sqrt{M}}\SOMMA{\mu=1}{K}\left(\dfrac{\a}{\sqrt{N}}\SOMMA{i=1}{N}\xi_i^\mu\sigma_i+\dfrac{\c}{\sqrt{L}}\SOMMA{k=1}{L}\chi_k^\mu\phi_k\right)\eta_j^\mu\, ,
     \\\\
     h^\phi_k(\bm\sigma, \bm\tau) =\dfrac{1}{\sqrt{L}}\SOMMA{\mu=1}{K}\left(\dfrac{\b}{\sqrt{N}}\SOMMA{i=1}{N}\xi_i^\mu\sigma_i+\dfrac{\c}{\sqrt{M}}\SOMMA{j=1}{M}\eta_j^\mu\tau_j\right)\chi_k^\mu\, ,
\end{array}
\end{equation}
in such a way that the Hamiltonian \eqref{H-TAM-diretta} can be recast as
\begin{equation}
 \mathcal{H}_{N,M,L}(\bm\sigma,\bm\tau,\bm\phi|\bm\xi,\bm\eta,\bm\chi) = - \SOMMA{i=1}{N}h_i^{\sigma}(\bm\tau, \bm\phi)\sigma_i  - \SOMMA{j=1}{M} h_j^{\tau}(\bm \sigma,\bm\phi)\tau_j  - \SOMMA{k=1}{L}h_k^{\phi}(\bm \sigma,\bm\tau)\phi_k
\end{equation}
and the noiseless evolutionary dynamics, obtained by setting $\beta \to \infty$ in the stochastic process \eqref{dinamical}, reads as \cite{Amit,Coolen} 
\begin{eqnarray}
\label{eq:evolv}
\sigma_i(t+1) &=& \textrm{sign} [h^\sigma_i(\bm\tau, \bm\phi, t)] \\
h^\sigma_i(t|\bm\tau, \bm\phi) &=& \dfrac{1}{\sqrt{N}}\SOMMA{\mu=1}{K}\left(\dfrac{\a}{\sqrt{M}}\SOMMA{j=1}{M}\eta_j^\mu\tau_j(t)+\dfrac{\b}{\sqrt{L}}\SOMMA{k=1}{L}\chi_k^\mu\phi_k(t)\right)\xi_i^\mu
\end{eqnarray}
where $t$ denotes the (discrete) time and $\bm h^\sigma \in \mathbb R^N$ is the local field acting on neurons in the $\sigma$-layer (stemming from the interactions with other neurons); for the other two layers, the dynamics is ruled by the same kind of equation, we only need to switch $\bm h^{\sigma}\to \bm h^{\tau}, \bm h^{\phi}$ and $\bm\sigma\to\bm\tau,\bm\phi$. 

We now investigate the one-step of neural dynamics using as a Cauchy datum a linear superposition of one pattern per dataset, that is we start the network in the mixture configuration $\bm \zeta \equiv \mathrm{sgn}\left[ \bm\xi^1+\bm\eta^1+\bm\chi^1\right]$.
In other words, we set $\bm\sigma(0) = \bm \tau (0) = \bm \phi (0) =\bm\zeta$.


It is more useful to write the neural dynamics as an evolutionary rule for the Mattis magnetizations introduced hereafter 
\begin{equation} 
\label{eq:magn_evolv}
\begin{array}{lll}
     m_{x^\mu}^{\sigma}(t+1):=\dfrac{1}{N}\SOMMA{i=1}{N}x_i^\mu \sigma_i(t+1)=\dfrac{1}{N}\SOMMA{i=1}{N}x_i^\mu \sigma_i(t)\mathrm{sign}\left[h^\sigma_i(\bm\tau, \bm\phi, t) \sigma_i(t)\right] ,
     \\\\
     m_{x^\mu}^{\tau}(t+1):=\dfrac{1}{M}\SOMMA{i=1}{M}x_i^\mu \tau_i(t+1)=\dfrac{1}{M}\SOMMA{i=1}{M}x_i^\mu \tau_i(t)\mathrm{sign}\left[h^\tau_i(\bm\sigma, \bm\phi, t) \tau_i(t)\right] ,
     \\\\
     m_{x^\mu}^{\phi}(t+1):=\dfrac{1}{L}\SOMMA{i=1}{L}x_i^\mu \phi_i(t+1)=\dfrac{1}{L}\SOMMA{i=1}{L}x_i^\mu \phi_i(t)\mathrm{sign}\left[h^\phi_i(\bm\tau, \bm\sigma, t) \phi_i(t)\right].
\end{array}
\end{equation}
where $\bm x^{\mu}\in(\bm\xi^\mu,\bm\eta^\mu, \bm\chi^\mu)$ and inspect how these magnetizations change from $t=0$ to $t=1$\footnote{The fact that pattern disentanglement can happen in just one step simply means that this neural architecture is optimal for such a task.}.
In particular, starting from an initial configuration $\bm\sigma(t=0)\equiv\bm\tau(t=0)\equiv\bm\phi(t=0)\equiv \bm\zeta \equiv \mathrm{sgn}\left[ \bm\xi^1+\bm\eta^1+\bm\chi^1\right]$, we want to check if the squared $3 \times 3$ matrix, whose entries are 
$$
\begin{bmatrix} 
m^\sigma_{\xi^1} & m^\sigma_{\eta^1} & m^\sigma_{\chi^1} \\
m^\tau_{\xi^1} & m^\tau_{\eta^1} & m^\tau_{\chi^1}\\
m^\phi_{\xi^1} & m^\phi_{\eta^1} & m^\phi_{\chi^1}. \\
\end{bmatrix}
\quad
$$
becomes diagonal, meaning that each layer retrieves the pattern pertaining to those it is handling and thus the initial mixture has been disentangled. 

In order to do that, let us focus on the $\sigma$-layer magnetization with respect to $\bm\xi^1$, assuming that $(\sigma(t=0), \bm\tau(t=0),\bm\phi(t=0))\equiv(\bm\zeta,\bm\zeta,\bm\zeta)$
\begin{equation}
    m_{\xi^1}^{\sigma}(1)=\dfrac{1}{N}\SOMMA{i=1}{N}\xi_i^1 \sigma_i(1)=\dfrac{1}{N}\SOMMA{i=1}{N}\xi_i^1 \zeta_i\mathrm{sign}\left[h^\sigma_i(\bm \zeta, \bm \zeta, t=0) \zeta_i\right],
    \label{eq:MC_updating_rule}
\end{equation}
using the relation $\bm\zeta = (\bm\xi^1+\bm\eta^1+\bm\chi^1-\bm \xi^1\bm\eta^1\bm\chi^1)/2$, we get
\begin{equation}
\begin{array}{lll}
     m_{\xi^1}^{\sigma}(1)&=&\dfrac{1}{N}\SOMMA{i=1}{N}\mathrm{sign}\left[h^\sigma_i(\bm \zeta, \bm \zeta, t) \left(\bm\xi^1+\bm\eta^1+\bm\chi^1-\bm \xi^1\bm\eta^1\bm\chi^1\right)\right] .
\end{array}
    \label{eq:MC_updating_rule_1}
\end{equation}
In the thermodynamic limit $(N \to \infty)$, the argument of the sign function in the r.h.s. of Eq. \eqref{eq:MC_updating_rule_1} can be approximated, by the CLT, as $2\zeta_i h^\sigma_i \sim \mu_1^{(\sigma,\xi)} + z_i\sqrt{\mu_2^{(\sigma,\xi)}-(\mu_1^{(\sigma,\xi)})^2}$, where $z_i \sim \mathcal{N}(0, 1)$, and
\begin{equation}
    \begin{array}{lll}
         \mu_1^{(\sigma,\xi)} = \mathbb{E}_{\bm\xi,\bm\eta,\bm\chi}\left[h^{\sigma}_i(\bm\zeta, \bm\zeta,t=0)\zeta_i\right]\,, &&\mu_2^{(\sigma,\xi)} = \mathbb{E}_{\bm\xi,\bm\eta,\bm\chi}\left[\Big(h^\sigma_i(\bm\zeta, \bm\zeta,t=0)\Big)^2\right].
    \end{array}
\end{equation}
Thus \eqref{eq:MC_updating_rule_1} becomes 
\begin{equation}
\begin{array}{lll}
     m_{\xi^1}^{\sigma}(1)&=&\dfrac{1}{N}\SOMMA{i=1}{N}\mathrm{sign}\left[\mu_1^{(\sigma,\xi)} + z_i\sqrt{\mu_2^{(\sigma,\xi)}-(\mu_1^{(\sigma,\xi)})^2}\right],
     \label{eq:magn_app1}
\end{array}
\end{equation}
then, recalling that
\begin{equation}
    \dfrac{1}{N}\SOMMA{i=1}{N}f(z_i) \xrightarrow[]{N\gg 1}\mathbb{E}_z f(z)
    \label{eq:magn_app2}
\end{equation}
and
\begin{equation}
    \mathbb{E}_z\mathrm{sign}\left(\mu_1+z \sqrt{\mu_2-\mu_1^2}\right)=\displaystyle\int\limits_{-\infty}^{+\infty}\dfrac{dz e^{-z^2/2}}{\sqrt{2\pi}}\mathrm{sign}\left(\mu_1+z \sqrt{\mu_2-\mu_1^2}\right) = \mathrm{erf}\left(\dfrac{\mu_1}{\sqrt{2(\mu_2-\mu_1^2)}}\right),
    \label{eq:magn_app3}
\end{equation}
after some calculations, we arrive to the following expression for the matrix of Mattis magnetizations:
\begin{equation}\label{Diagonale}
\begin{bmatrix} 
 m^\sigma_{\xi^1}=\mathrm{erf}\left[\dfrac{1}{\alpha\theta}\dfrac{a\alpha+b\theta}{\sqrt{2(a^2+b^2)\gamma}}\right] & m^\sigma_{\eta^1}=0 & m^\sigma_{\chi^1}=0 \\
 m^\tau_{\xi^1}=0 & m^\tau_{\eta^1}=\mathrm{erf}\left[\dfrac{\theta}{\alpha}\dfrac{\alpha a+c}{\sqrt{2(a^2+c^2)\gamma}}\right] & m^\tau_{\chi^1}=0\\
 m^\phi_{\xi^1}=0 & m^\phi_{\eta^1}=0 & m^\phi_{\chi^1}=\mathrm{erf}\left[\dfrac{\alpha}{\theta}\dfrac{\theta b+c}{\sqrt{2(b^2+c^2)\gamma}}\right] \\
\end{bmatrix}
\quad
\end{equation}
As expected, the 3-pattern mixture used as input signal has been disentangled in its basic components, i.e. the pattern themselves.

\subsection{Homogeneous mixtures}
\label{app:S2N_0}
Under the working assumption of Sec.\ref{Sec:Penalty}, following the same steps presented in Sec.\ref{Hinton}, we define the updating rules which govern the noiseless neural dynamics of Hamiltonian \eqref{eq:dis_hamiltonian_magn} as: 
\begin{eqnarray}
\label{eq:evolv_tau}
\tau_i(t+1) &=& \tau_i(t)\textrm{sign} \Big[ h^\tau_i(t|\bm\sigma\equiv\bm\xi^{(1,2)}, \bm\phi)\tau_i(t)\Big] \\
 h^\tau_i(t|\bm\sigma\equiv\bm\xi^{(1,2)},\bm\phi) &=& \dfrac{1}{N}\SOMMA{\mu=1}{K}\left(\SOMMA{j=1}{N}\xi_j^\mu\xi^{(1,2)}_j-|\c|\SOMMA{k=1}{N}\xi_k^\mu\phi_k\right)\xi_i^\mu.
\end{eqnarray}
The dynamics of the other layers follow the same form, with the substitutions $\bm h^{\tau} \to \bm h^{\phi}$ and $\bm\tau \to \bm\phi$.

As done in Sec.\ref{Hinton} is more practical to express the neural dynamics in terms of the evolution of the Mattis magnetizations: 
\begin{equation} 
\label{eq:magn_evolv_tau_phi}
\begin{array}{lll}
     m_{\xi^\mu}^{\tau}(t+1):=\dfrac{1}{N}\SOMMA{i=1}{N}\xi_i^\mu \tau_i(t+1)=\dfrac{1}{N}\SOMMA{i=1}{N}\xi_i^\mu \tau_i(t)\mathrm{sign}\left[h^\tau_i(t|\bm\sigma\equiv\bm\xi^{(1,2)},\bm\phi) \tau_i(t)\right] ,
     \\\\
     m_{\xi^\mu}^{\phi}(t+1):=\dfrac{1}{N}\SOMMA{i=1}{N}\xi_i^\mu \phi_i(t+1)=\dfrac{1}{N}\SOMMA{i=1}{N}\xi_i^\mu \phi_i(t)\mathrm{sign}\left[h^\phi_i(t|\bm\sigma\equiv\bm\xi^{(1,2)},\bm\tau) \phi_i(t)\right].
\end{array}
\end{equation}
We now analyze how these magnetizations evolve from $t=0$ to $t=1$. Specifically, we focus on the first step of the neural dynamics, with the input layer clamped to $\bm \xi^{(1,2)} \equiv \mathrm{sign}\left[ \bm\xi^1 + \bm\xi^2 \right]$. Thus, we initialize $\bm\sigma(0) = \bm \xi^{(1,2)}$ and investigate the behavior of the magnetizations for the $\tau$ and $\phi$ layers with respect to $\bm\xi^1$, $\bm\xi^2$ and $\bm\xi^{\nu>2}$ across three initial configurations:

\begin{itemize}
    \item $( \bm\tau(0),\bm\phi(0))\equiv(\bm\xi^1,\bm\xi^1)$: output layers aligned to the same component of the input layer
    \begin{equation}
        \begin{array}{lll}
             m_{\xi^1}^{\tau}(1)=\dfrac{1}{N}\SOMMA{i=1}{N}\mathrm{sign}\left[h^\tau_i(0|\bm\xi^{(1,2)},\bm\xi^1) \xi^1_i\right],&&m_{\xi^1}^{\phi}(1)=\dfrac{1}{N}\SOMMA{i=1}{N}\mathrm{sign}\left[h^\phi_i(0|\bm\xi^{(1,2)},\bm\xi^1) \xi^1_i\right],
             \\\\
             m_{\xi^2}^{\tau}(1)=\dfrac{1}{N}\SOMMA{i=1}{N}\mathrm{sign}\left[h^\tau_i(0|\bm\xi^{(1,2)},\bm\xi^1) \xi^2_i\right],&&m_{\xi^2}^{\phi}(1)=\dfrac{1}{N}\SOMMA{i=1}{N}\mathrm{sign}\left[h^\phi_i(0|\bm\xi^{(1,2)},\bm\xi^1) \xi^2_i\right],
             \\\\
             m_{\xi^{\nu>2}}^{\tau}(1)=\dfrac{1}{N}\SOMMA{i=1}{N}\mathrm{sign}\left[h^\tau_i(0|\bm\xi^{(1,2)},\bm\xi^1) \xi^\nu_i\right],&&m_{\xi^{\nu>2}}^{\phi}(1)=\dfrac{1}{N}\SOMMA{i=1}{N}\mathrm{sign}\left[h^\phi_i(0|\bm\xi^{(1,2)},\bm\xi^1) \xi^\nu_i\right],
        \end{array}
        \label{eq:111}
    \end{equation}
    \item $( \bm\tau(0),\bm\phi(0))\equiv(\bm\xi^1,\bm\xi^2)$: output layers aligned to different components of the input layer
    \begin{equation}
        \begin{array}{lll}
             m_{\xi^1}^{\tau}(1)=\dfrac{1}{N}\SOMMA{i=1}{N}\mathrm{sign}\left[h^\tau_i(0|\bm\xi^{(1,2)},\bm\xi^2) \xi^1_i\right],&&m_{\xi^1}^{\phi}(1)=\dfrac{1}{N}\SOMMA{i=1}{N}\mathrm{sign}\left[h^\phi_i(0|\bm\xi^{(1,2)},\bm\xi^1) \xi^1_i\right],
             \\\\
             m_{\xi^2}^{\tau}(1)=\dfrac{1}{N}\SOMMA{i=1}{N}\mathrm{sign}\left[h^\tau_i(0|\bm\xi^{(1,2)},\bm\xi^2) \xi^2_i\right],&&m_{\xi^2}^{\phi}(1)=\dfrac{1}{N}\SOMMA{i=1}{N}\mathrm{sign}\left[h^\phi_i(0|\bm\xi^{(1,2)},\bm\xi^1) \xi^2_i\right],
             \\\\
             m_{\xi^{\nu>2}}^{\tau}(1)=\dfrac{1}{N}\SOMMA{i=1}{N}\mathrm{sign}\left[h^\tau_i(0|\bm\xi^{(1,2)},\bm\xi^2) \xi^\nu_i\right],&&m_{\xi^{\nu>2}}^{\phi}(1)=\dfrac{1}{N}\SOMMA{i=1}{N}\mathrm{sign}\left[h^\phi_i(0|\bm\xi^{(1,2)},\bm\xi^1) \xi^\nu_i\right],
        \end{array}
        \label{eq:112}
    \end{equation}
    \item $(\bm\tau(0),\bm\phi(0))\equiv(\bm\xi^1,\bm\xi^{\nu>2})$: only output layers aligned to one of the  components of the input layer
    \begin{equation}
        \begin{array}{lll}
             m_{\xi^1}^{\tau}(1)=\dfrac{1}{N}\SOMMA{i=1}{N}\mathrm{sign}\left[h^\tau_i(0|\bm\xi^{(1,2)},\bm\xi^\nu) \xi^1_i\right],&&m_{\xi^1}^{\phi}(1)=\dfrac{1}{N}\SOMMA{i=1}{N}\mathrm{sign}\left[h^\phi_i(0|\bm\xi^{(1,2)},\bm\xi^1) \xi^1_i\right],
             \\\\
             m_{\xi^2}^{\tau}(1)=\dfrac{1}{N}\SOMMA{i=1}{N}\mathrm{sign}\left[h^\tau_i(0|\bm\xi^{(1,2)},\bm\xi^\nu) \xi^2_i\right],&&m_{\xi^2}^{\phi}(1)=\dfrac{1}{N}\SOMMA{i=1}{N}\mathrm{sign}\left[h^\phi_i(0|\bm\xi^{(1,2)},\bm\xi^1) \xi^2_i\right],
             \\\\
             m_{\xi^{\nu>2}}^{\tau}(1)=\dfrac{1}{N}\SOMMA{i=1}{N}\mathrm{sign}\left[h^\tau_i(0|\bm\xi^{(1,2)},\bm\xi^\nu) \xi^\nu_i\right],&&m_{\xi^{\nu>2}}^{\phi}(1)=\dfrac{1}{N}\SOMMA{i=1}{N}\mathrm{sign}\left[h^\phi_i(0|\bm\xi^{(1,2)},\bm\xi^1) \xi^\nu_i\right],
        \end{array}
        \label{eq:113}
    \end{equation}
\end{itemize}

In the thermodynamic limit $(N \to \infty)$, the argument of the sign functions in the r.h.s. of Eqs.~\eqref{eq:111}~-~\eqref{eq:112}~-~\eqref{eq:113} can be approximated, by the CLT, as 
$$ h^{\bm y_1}_i(0|\bm\xi^{(1,2)},\bm  y_2) x_i\sim \mu_1^{(\bm  y_2,\bm x)} (\bm y_1)+ z_i\sqrt{\mu_2^{(\bm  y_2,\bm x)}(\bm y_1)-(\mu_1^{(\bm  y_2,\bm x)}(\bm y_1))^2},$$
where $z_i \sim \mathcal{N}(0, 1)$, $\bm{x}\in\{\bm\xi^1,\bm\xi^2,\bm\xi^{\nu}\}$, $\bm y_1 \in \{\bm\tau, \bm\phi\}$,  $\bm y_2 \in \{\bm\xi^1, \bm\xi^2, \bm\xi^{\nu>2}\}$ and
\begin{equation}
    \begin{array}{lll}
         \mu_1^{(\bm  y_2,\bm{x})}(\bm y_1) = \mathbb{E}_{\bm\xi}\left[h^{ \bm y_2}_i(0|\bm\xi^{(1,2)}, \bm  y_2)x_i\right]\,, &&\mu_2^{(\bm y_2,\bm x)} (\bm y_1)= \mathbb{E}_{\bm\xi}\left[\Big(h^{\bm y_1}_i(0|\bm\xi^{(1,2)}, \bm  y_2)\Big)^2\right].
    \end{array}
\end{equation}

Thus, Eqs. \eqref{eq:111}, \eqref{eq:112}, and \eqref{eq:113} can be reformulated following the same steps presented previously in Eqs. \eqref{eq:magn_app1}--\eqref{eq:magn_app2}--\eqref{eq:magn_app3} as follows:
\begin{equation}
    \begin{array}{lll}
         m_{\bm x}^{ \bm y_1}(1)&\xrightarrow[]{N\gg 1}&\mathrm{erf}\left(\dfrac{\mu_1^{(  y_2,\bm x)}(\bm y_1)}{\sqrt{2\left(\mu_2^{(  y_2,\bm x)}(\bm y_1)-(\mu_1^{(  y_2,\bm x)}(\bm y_1))^2\right)}}\right)
    \end{array}
    \label{eq:erf_Mattis}
\end{equation}
which represents the alignment of the layer $\bm y_1 \in \{\bm\tau, \bm\phi\}$ with respect to the pattern $\bm x \in \{\bm\xi^1, \bm\xi^2, \bm\xi^{\nu>2}\}$,
evaluated at the first time step t = 1, being $\bm y_2 \in \{\bm\xi^1, \bm\xi^2, \bm\xi^{\nu>2}\}$ the initial setting of the layer opposite to $y_1$ (in fact, when updating the layer $\tau$ we need to specify only the state of the layer $\phi$, and vice versa, because the layer $\sigma$ is clamped).

\subsection{Biased patterns} \label{sec:S2N_biased}
Working under the constraint of Sec.\ref{bias}, following the same steps presented in the previous Sections, we will analyze the evolution of Mattis magnetization
after one step of neural dynamics.

Let define the internal field acting on each neuron and the related update rule of Mattis magnetization
\begin{equation}
\begin{array}{lll}
    \phi_i^{(n+1)}&=&\phi_i^{(n)}\mathrm{sign}\left[ h_i(\bm\xi^{(1,2)},\bm\tau^{(n)})\phi_i^{(n)}\right]
   \\\\
   &=&\phi_i^{(n)}\mathrm{sign}\left[\dfrac{1}{N}\SOMMA{\mu}{K}\SOMMA{j}{N}\xi_j^{\mu}\xi^{(1,2)}_j\xi_i^{\mu}\phi_i^{(n)}-\dfrac{|\c|}{N}\SOMMA{\mu}{K}\SOMMA{j}{N}\xi_{j}^{\mu}\tau_{j}^{(n)}\xi_i^{\mu}\phi_i^{(n)} - g  \dfrac{1}{N}\SOMMA{j}{N}\phi_j^{(n)}\phi_i^{(n)}- 2 g b \phi_i^{(n)}\right]
   \\\\
   m_\mu^{(n+1)}(\bm\phi)&=&\dfrac{1}{N}\SOMMA{i=1}{N}\xi_i^{\mu}\phi_i^{(n+1)}=\dfrac{1}{N}\SOMMA{i=1}{N}\xi_i^{\mu}\phi_i^{(n)}\mathrm{sign}\left[ h_i(\bm\xi^{(1,2)},\bm\tau^{(n)})\phi_i^{(n)}\right],
    \label{eq:update_mattis}
\end{array}
\end{equation}
we now study, as done previously for the unbiased case, the behaviour of the Mattis magnetization after one step of neuronal updating: we  analyze three cases and for all the cases we compute the average value of the argument of the $\mathrm{sign}$ function as follows.
\begin{itemize}
    \item $(\bm\tau(0),\bm\phi(0))\equiv(\bm\xi^1,\bm\xi^1)$
    \begin{equation}
        \begin{array}{lllll}
             \mu_1^{(\bm\xi^1,\bm\xi^1)}(\bm \tau)=\dfrac{(1+b^2)^2}{2}+(K-2)b^4+ g b^2-|\c|\left(1+(K-1)b^4\right)
             \\\\
             \mu_1^{(\bm\xi^1,\bm\xi^1)}(\bm\phi)=\dfrac{(1+b^2)^2}{2}+(K-2)b^4+ g b^2-|\c|\left(1+(K-1)b^4\right)
        \end{array}
        \label{eq:111_1_b}
    \end{equation}
    \item $( \bm\tau(0),\bm\phi(0))\equiv(\bm\xi^1,\bm\xi^2)$
    \begin{equation}
        \begin{array}{lllll}
             \mu_1^{(\bm\xi^2,\bm\xi^1)}(\bm\tau)=\dfrac{(1+b^2)^2}{2}+(K-2)b^4+ g b^2-|\c| \left(2b^2+  (K-2)b^4\right),
             \\\\
             \mu_1^{(\bm\xi^2,\bm\xi^1)}(\bm\phi)=\dfrac{(1+b^2)^2}{2}+(K-2)b^4+ g b^2-|\c| \left(2b^2+  (K-2)b^4\right),
        \end{array}
        \label{eq:112_1_b}
    \end{equation}   
    \item $(\bm\tau(0),\bm\phi(0))\equiv(\bm\xi^1,\bm\xi^{\nu>2})$
    \begin{equation}
        \begin{array}{lllll}
             \mu_1^{(\bm\xi^\nu,\bm\xi^1)}(\bm\tau)=\dfrac{(1+b^2)^2}{2}+(K-2)b^4+ g b^2-|\c| \left(2b^2+  (K-2)b^4\right),
             \\\\
             \mu_1^{(\bm\xi^\nu,\bm\xi^1)}(\bm\phi)=2b^2+(K-2)b^4+ g b^2-|\c| \left(2b^2+  (K-2)b^4\right).
        \end{array}
        \label{eq:113_1_b}
    \end{equation}
\end{itemize}
In order to have the best separation possible between the two outputs, we can tune the value of $|\c|$ to set the mean value of the first case (both layers aligned) equal to the last line of the third case (one output disaligned with the input), namely
\begin{equation}
    \begin{array}{lll}
    \mu_1^{(\bm\xi^1,\bm\xi^1)}(\bm \tau)&=&\mu_1^{(\bm\xi^\nu,\bm\xi^1)}(\bm\tau) 
    \end{array}
\end{equation}
this equivalence gives rise to the following constraint
\begin{equation}
    |\c*|=\dfrac{1}{2}.
\end{equation}
Fixing $\c$ in this way guarantees that the maximal separation among the output layers\footnote{We stress that in this case, as a bias is present in the network, two layers at least must have an average value of the overlap equal to $b^2$.}, note further that $|\c^*|$ is $K$ and $b$-independent.

\end{document}